\documentclass[aps,prd,amsmath,floats,floatfix, twocolumn,
superscriptaddress,nofootinbib,showpacs]{revtex4-1}

\usepackage{ulem}
\usepackage{array,mathtools,amssymb,booktabs}
\usepackage{siunitx}
\usepackage{graphicx}
\usepackage[export]{adjustbox}
\usepackage[usenames,dvipsnames]{xcolor}
\usepackage[colorlinks, pdfborder={0 0 0}, plainpages=false]{hyperref}

\newcolumntype{C}{>{$}c<{$}}
\AtBeginDocument{
\heavyrulewidth=.08em
\lightrulewidth=.05em
\cmidrulewidth=.03em
\belowrulesep=.65ex
\belowbottomsep=0pt
\aboverulesep=.4ex
\abovetopsep=0pt
\cmidrulesep=\doublerulesep
\cmidrulekern=.5em
\defaultaddspace=.5em
}

\begin{document}

\newcommand{\Caltech}{\affiliation{TAPIR, Walter Burke Institute for Theoretical Physics, MC 350-17,
    California Institute of Technology, Pasadena, California 91125, USA}}
\newcommand{\Einstein}{\affiliation{NASA Einstein Fellow}}
\newcommand{\Cornell}{\affiliation{Cornell Center for Astrophysics and
    Planetary Science, Cornell University, Ithaca, New York, 14853, USA}}
\newcommand{\WSU}{\affiliation{Department of Physics \& Astronomy,
	Washington State University, Pullman, Washington 99164, USA}}
\newcommand{\CITA}{\affiliation{Canadian Institute for Theoretical 
    Astrophysics, University of Toronto, Toronto, Ontario M5S 3H8, Canada}}
\newcommand{\UofT}{\affiliation{Department of Physics,
    University of Toronto, Toronto, Ontario, M5S 3H5, Canada}}
\newcommand{\CIFAR}{\affiliation{Canadian Institute for Advanced Research, 180 Dundas St.~West, Toronto, ON M5G 1Z8, Canada}} %
\newcommand{\LBL}{\affiliation{Nuclear Science Division, Lawrence Berkeley National Laboratory,
1 Cyclotron Rd, Berkeley, CA 94720, USA}}
\newcommand{\NCSU}{\affiliation{Department of Physics, North Carolina State University, Raleigh, North Carolina 27695, USA; Hubble Fellow}}
\newcommand{\AEI}{\affiliation{Max Planck Institute for Gravitational Physics (Albert Einstein Institute), Am M\"uhlenberg 1, Potsdam 14476, Germany}}
\newcommand{\Maryland}{\affiliation{Department of Physics, University of Maryland, College Park, MD 20742, USA}}
\newcommand{\UCBAstro}{\affiliation{Astronomy Department and Theoretical Astrophysics Center, 
University of California, Berkeley, 601 Campbell Hall, Berkeley CA, 94720}}
\newcommand{\UCBPhysics}{\affiliation{Department of Physics, University of California, Berkeley, Le Conte Hall, Berkeley, CA 94720}}
\newcommand{\UNH}{\affiliation {Department of Physics, University of New Hampshire, 9 Library Way, Durham NH 03824, USA}}
\newcommand{\NCSA}{\affiliation{NCSA, University of Illinois at Urbana-Champaign, Urbana, Illinois, 61801, USA}} %
\newcommand{\RU}{\affiliation{Department of Astrophysics/IMAPP, Radboud University Nijmegen, P.O. Box 9010, 6500 GL Nijmegen, The Netherlands}}
\newcommand{\GRAPPA}{\affiliation{GRAPPA, Anton Pannekoek Institute for Astronomy and Institute of High-Energy Physics, University of Amsterdam, Science Park 904, 1098 XH Amsterdam, The Netherlands}}
\newcommand{\DeltaITP}{\affiliation{Delta Institute for Theoretical Physics, Science Park 904, 1090 GL Amsterdam, The Netherlands}}
\newcommand{\Nikhef}{\affiliation{Nikhef, Science Park 105, 1098 XG Amsterdam, The Netherlands}}

\title{
Unequal Mass Binary Neutron Star Simulations with Neutrino Transport: \\ Ejecta and Neutrino Emission
}

\author{Trevor Vincent}\CITA\UofT
\author{Francois Foucart}\UNH
\author{Matthew D. Duez}\WSU
\author{Roland Haas}\NCSA
\author{Lawrence E. Kidder}\Cornell
\author{Harald P. Pfeiffer}\AEI\CITA
\author{Mark A. Scheel}\Caltech

\begin{abstract}
We present twelve new simulations of unequal mass neutron star mergers. The simulations were preformed with the SpEC code, and utilize nuclear-theory based equations of state and a two-moment gray neutrino transport scheme with an improved energy estimate based on evolving the number density. We model the neutron stars with the SFHo, LS220 and DD2 equations of state (EOS) and we study the neutrino and matter emission of all twelve models to search for robust trends between binary parameters and emission characteristics. We find that the total mass of the dynamical ejecta exceeds $0.01M_\odot$ only for SFHo with weak dependence on the mass-ratio across all models.  We find that the  ejecta have a broad electron fraction ($Y_e$) distribution ($\approx 0.06-0.48$), with mean $0.2$. $Y_e$ increases with neutrino irradiation over time, but decreases with increasing binary asymmetry. We also find that the models have ejecta with a broad asymptotic velocity distribution ($\approx 0.05-0.7c$). The average velocity lies in the range $0.2c - 0.3c$ and decreases with binary asymmetry. Furthermore, we find that disk mass increases with binary asymmetry and stiffness of the EOS. The $Y_e$ of the disk increases with softness of the EOS. The strongest neutrino emission occurs for the models with soft EOS. For (anti) electron neutrinos we find no significant dependence of the magnitude or angular distribution or neutrino luminosity with mass-ratio. The heavier neutrino species have a luminosity dependence on mass-ratio but an angular distribution which does not change with mass-ratio.
\end{abstract}

\maketitle

%%%%%%%%%%%%%%%%%%%%%%%%%%%%%%%%%%%%%%%%%%%%%%%%%%%%%%%%%%%%%%%%%%%%%%%%%%%%%%%
\section{Introduction}

%%%%%%%%%%%%%%%%%%%%%%%%%%%%%%%%%%%%%%%%%%%%%%%%%%%%%%%%%%%%%%%%%%%%%%%%%%%%%%%
The binary NS merger GW170817 was a landmark event, combining the first detection of gravitational waves
with an observation of a short gamma ray burst and a kilonova~\cite{TheLIGOScientific:2017qsa,GBM:2017lvd,2017ApJ...848L..13A,ajello2018fermi}. Following the event, studies emerged analyzing many aspects of GW170817, from the
internal structure of neutron stars~\cite{gw170817-pe}, to the production of short gamma-ray bursts~\cite{moch:93,Lee1999a,Janka1999,GBM:2017lvd,2017ApJ...848L..13A,2018Natur.561..355M} and the synthesis of r-process elements~\cite{Li:1998bw,1976ApJ...210..549L,Rosswog:1998hy, 2005astro.ph.10256K,2010MNRAS.406.2650M,metzger2017,2017Sci...358.1559K,2017Sci...358.1556C,2017ApJ...848L..19C,2017Sci...358.1556C,Cowperthwaite:2017dyu,2017Natur.551...80K,2017Sci...358.1583K,2017ApJ...848L..32M,2017ApJ...848L..18N,2017Natur.551...67P,2017Natur.551...75S,2017ApJ...848L..16S,2017ApJ...848L..27T,2017Sci...358.1565E}. With the increasing sensitivity of advanced gravitational wave
interferometers many binary neutron star (BNS) detections are expected in the next decade \cite{ligo2018gwtc}.

Numerical simulations of mergers
play a crucial role in efforts to model the gravitational
wave signal, predict the properties of its electromagnetic counterparts,
and estimate the production of r-process elements from the merger. In this work, we
focus on the matter and neutrino emissions from BNS mergers. In particular we look at the ejecta
and neutrino emission from a new set of twelve BNS simulations
which extends a previous set of four equal mass BNS simulations \cite{foucart:2015gaa} to include mass-ratios not equal to one. While asymmetric mass-ratio
BNS simulations have been studied before in the context of general-relativistic-radiation
hydrodynamics, e.g. \cite{sekiguchi2016dynamical}, \cite{lehner2016unequal}, \cite{radice2016dynamical},
these previous studies use either a simpler neutrino scheme and/or
different mass-ratios and EOS. Thus this paper adds to the ongoing effort to simulate
BNS systems and characterize their observables.

Neutrino interactions were first included in general relativistic simulations of neutron star mergers through a simple leakage scheme \cite{sekiguchi:2011zd}, based on approximate methods developed for Newtonian simulations \cite{ruffert1996,rosswog:2003rv}. A leakage scheme uses the local properties of the fluid and an estimate of the neutrino optical depth to determine the amount of energy lost locally to neutrino-matter interactions, and the associated change in the composition of the fluid. Leakage schemes provide an order-of-magnitude accurate estimate of neutrino cooling in the post-merger remnant, and have thus been used to capture the first-order effect of neutrino-matter interactions in general relativistic simulations of compact binary mergers \cite{sekiguchi:2011zd,wanajo2014,lehner2016unequal,radice2016dynamical,palenzuela2015,deaton2013black,foucart2014neutron}. The inclusion of neutrino-matter effects with the leakage scheme, while crude, significantly affected the composition, morphology and total mass of the outflows, with some studies showing a factor 2 difference in total ejecta mass \cite{radice2016dynamical}. However, most implementations of leakage do not account for irradiation of low density regions by neutrinos emitted from hot, dense regions. This potentially leads to large errors in the composition of the outflows, mostly by underestimating the number of protons \cite{foucartm1:2016,foucart2015post}. Accordingly, the simplest leakage schemes are very inaccurate when attempting to predict the properties of post-merger electromagnetic signals. The only general relativistic simulations going beyond leakage use a moment formalism with an analytic closure to approximate the Boltzmann equation \cite{1981mnras.194..439t,shibata:11}. In particular, neutron star merger simulations have been performed with a gray two-moment scheme \cite{foucartm1:2016,foucart2015post,sekiguchi2015dynamical,sekiguchi2016dynamical}, in which the energy density and flux density of each neutrino species are evolved. In BNS mergers, the use of this moment formalism showed that a range of compositions and thus of nucleosynthesis outcomes, exists in the material ejected by the merger \cite{wanajo2014}. 

Previous studies of asymmetric mass-ratio BNS systems with fully general-relativistic radiation-hydrodynamics include
\cite{sekiguchi2016dynamical,lehner2016unequal,radice2018binary}. Sekiguchi et al.~\cite{sekiguchi2016dynamical} studied two EOS, SFHo and DD2 with mass ratios between .86 and 1.0 and a fixed total mass of $2.7M_\odot$ to around 30-ms post-merger using a two-moment neutrino transport scheme. They found that for SFHo the ejecta mass depended weakly on mass-ratio, but the average electron number per baryon decreased with mass ratio. For DD2 these trends were reversed. They also found that only the soft EOS, SFHo, produced ejecta mass above $0.01M_\odot$. Lehner et al.~\cite{lehner2016unequal} studied three EOS, NL3, SFHo and DD2 with mass ratios between .76 and 1.0 and a fixed total mass of $2.7M_\odot$ at 3-ms post-merger using a neutrino leakage scheme. They found that there was a greater ejecta mass with increasing binary asymmetry. Unlike Sekiguchi et al., Lehner et. al found that none of the EOS produced ejecta above $0.01M_\odot$. Finally, Radice et al. \cite{radice2018binary} studied four EOS, BHB$\Lambda\phi$, SFHo, DD2 and LS220, with mass ratios between .85 and 1.0 and a fixed total mass of $2.7M_\odot$ to around 20-ms post-merger  using a neutrino leakage scheme and a viscous hydrodynamics scheme. Radice et al./ found that  none of their models produced ejecta over $0.01M_\odot$ and their numbers agreed with Lehner et al.~\cite{lehner2016unequal}. The discrepancy between these results could however be due different choices for the definition of the unbound material in these studies. This paper adds onto previous works in the following ways. First, we look at a different set of parameters not found in the above studies. We use the SFHo, LS220 and DD2 EOS to study the effects of EOS on the merger emissions. We use mass-ratios ranging from $q \sim.76-1$ to study the effects of mass asymmetry on emissions. Unlike the previous studies, we do not fix the total mass and allow it to vary from $\sim 2.5-2.9M_\odot$. On top of this, we use a new two-moment neutrino transport scheme which evolves the number density, allowing for consistent lepton number evolution \cite{foucart:2016rxm}. 

We organize the paper as follows. In Section~\ref{sec:num_imp}, we discuss the numerical implementation we use and the
equations we solve, including the gray two moment scheme for neutrino transport. In the following
sections of the paper, we discuss the matter and neutrino emission from a new set of
twelve binary neutron star merger simulations, ranging in mass ratio and equation of state.
Finally we conclude with ideas for future work. We use a system of units such
that $c = G = M_\odot = 1$, where c is the speed of light in
vacuum, G is the gravitational constant, and $M_\odot$ is the mass
of the Sun. We use Einstein's convention of summation over
repeated indices. Latin indices run over 1, 2, 3, while Greek
indices run over 0, 1, 2, 3. The spacetime metric signature is $(-,+,+,+)$.

\section{Numerical Implementation}
\label{sec:num_imp}
\subsection{General Overview}

We evolve Einstein's equations and the general relativistic equations of ideal radiation-hydrodynamics using the Spectral Einstein Code (SpEC)\cite{specwebsite}. SpEC evolves those equations on
two separate grids: a pseudospectral grid for Einstein's equations, written in the generalized harmonic formulation \cite{lindblom2006},
and a finite volume grid for the general relativistic equations
of neutrino-hydrodynamics, written in conservative form. The latter
makes use of an approximate Riemann solver (HLL \cite{hll})
and high-order shock capturing methods (fifth order WENO
scheme \cite{liu1994200,jiang1996202}), resulting in a second-order accurate evolution scheme. For the time evolution, we use a third-order
Runge-Kutta algorithm. Finally, after each time step, the two
grids communicate the required source terms, using a third-order accurate spatial interpolation scheme. Those source terms are the metric and its derivatives (from the pseudospectral grid to the finite volume grid) and the fluid variables, which are rest-mass density, pressure, spatial velocity, the Lorentz factor and enthalpy. The following sections will give more detail on individual segments of this numerical method.

In the following sections, we make use the 3+1 decomposition of the metric
\begin{align}
  ds^2 &= g_{\alpha\beta}dx^\alpha dx^\beta \\
  &= -\alpha^2dt^2 + \gamma_{ij}(dx^i + \beta^i)(dx^j + \beta^j)
\end{align}
where $\alpha$ is the lapse, $\beta^i$ the shift, and $\gamma_{ij}$ the 3-metric on a slice of constant coordinate t. The extension of $\gamma_{ij}$ to the full 4-dimensional space is the projection operator:
\begin{equation}
  \gamma_{\alpha\beta} = g_{\alpha\beta} + n_\alpha n_\beta,
\end{equation}
with $n_\mu=(-\alpha,0,0,0)$ the unit normal to a $t={\rm constant}$ slice.

\subsection{Initial Data}

Initial data for this simulation was produced by an BNS initial data solver based on the work of Foucart~{\it et
al}. for BHNS systems \cite{foucart2008initial}, which was built upon the elliptic solver Spells \cite{pfeiffer2003} and further improved for BNS systems in \cite{tacik2015binary} and \cite{haas:2016}. We start by considering systems in quasi-equilibrium, where time
derivatives vanish in a corotating frame. We take the metric to be conformally flat and solve for the lapse,
shift, and conformal factor using the extended conformal thin sandwich (XCTS) equations \cite{pfeiffer2003b}. The matter in the stars is modeled as a cold perfect fluid with an irrotational velocity profile.

Low eccentricity can be achieved through an iterative procedure requiring the evolution of the system for 2-3 orbits in each iteration \cite{pfeiffer2007reducing}. All of the simulations in this paper use that algorithm to achieve estimated eccentricities of approximately $e\sim 0.001$.

Neutrinos are initialized in thermal equilibrium with the fluid. As the initial data is composed of two cold neutron stars ($T=0.1{\rm MeV}$), the initial energy density of neutrinos is negligible compared to its value at any later time in the evolution.

\subsection{Spacetime Evolution}

The Einstein field equations can be written in the form
\begin{equation}
  \label{eqn:efe}
R_{\alpha\beta} = 8\pi\left(T_{\alpha\beta} - \frac{1}{2} g_{\alpha\beta}T\right),
\end{equation}
where $R_{\alpha\beta}$ is the Ricci tensor, $ g_{\alpha\beta}$ is the metric tensor and $T_{\alpha\beta}$ is the stress energy tensor with trace T. SpEC uses the generalized harmonic decomposition \cite{lindblom2006} to write the Einstein field equations in a form that allows stable numerical computation.

In the generalized harmonic formalism, the evolution of the coordinates follows the wave equation
\begin{equation}
  \label{eqn:gh_coordinates}
   g_{\alpha\beta}\nabla^\zeta\nabla_\zeta x^\beta = H_\alpha\left(x^\epsilon\right),
\end{equation}
where $ g_{ab}$ is the spacetime metric, and $H_\alpha(x^c)$ a set of four arbitrary functions. Using Eq.~(\ref{eqn:gh_coordinates}) we can rewrite 
Eq.~(\ref{eqn:efe}) as \cite{pretorius2005numerical}
\begin{equation}
  \begin{split}
  \label{eqn:efe_gh}
   g^{\delta\gamma}\partial_\gamma\partial_\delta g_{\alpha\beta} + \partial_\beta g^{\gamma\delta}\partial_\gamma g_{\alpha\delta} &+ \partial_\alpha g^{\gamma\delta}\partial_\gamma g_{\beta\delta} + 2\partial_{(\beta,} H_{\alpha)} \\
  - 2H_\delta\Gamma^\delta_{\alpha\beta} + 2\Gamma^\gamma_{\delta\beta}\Gamma^{\delta}_{\gamma\alpha} &= -8\pi(2T_{\alpha\beta} - g_{\alpha\beta}T),
  \end{split}
\end{equation}
where the Christoffel symbols $\Gamma^\delta_{\alpha\beta}$ are defined by
\begin{equation}
\Gamma^\gamma_{\alpha\beta} = \frac{1}{2} g^{\gamma\epsilon}\left(\partial_\beta g_{\alpha\epsilon} + \partial_\alpha g_{\beta\epsilon} - \partial_\epsilon g_{\alpha\beta}\right).
\end{equation}
Eq.~(\ref{eqn:efe_gh}) introduces four independent gauge functions $H_\alpha$, which need to be chosen. At t = 0, we set
\begin{equation}
H_\alpha\left(x^i_c,t\right) = H_\alpha\left(x^i_c,0\right)\exp\left(-\frac{t^2}{\tau^2}\right),
\end{equation}
with $\tau = \sqrt{d^3_0/M_\infty}$, $d_0$ the initial separation, $M_\infty$ is the total mass of the binary at infinite separation and $x^i_c$ are comoving spatial coordinates which follow the rotation and inspiral of the binary. The initial data is constructed in a gauge $H^{initial}_a$ that assumes the time derivatives in the comoving frame are zero. At the beginning of the simulation, we set $H_\alpha\left(x^i_c,0\right) = \Hat H_\alpha$, where $\Hat H_\alpha$ is a tensor that agrees with $H^{initial}_a$ in a frame comoving with the grid and is constant in time. The gauge will thus evolve into the harmonic condition $H_\alpha = 0$. During merger, we find that transitioning to the ``damped harmonic'' gauge condition leads to more accurate numerical evolution \cite{szilagyi2014key}. Defining
\begin{equation}
  H_\alpha = \left(\log{\frac{\sqrt{\gamma}}{\alpha}}\right)^2\left(\frac{\sqrt{\gamma}}{\alpha}t_\alpha - \gamma_{\alpha i} \frac{\beta^i}{\alpha}\right),
\end{equation}
we transition according to
\begin{equation}
  H_\alpha(t) = H_\alpha\left(1-\exp{\frac{-(t-t_{DH})^2}{\tau_{m}^2}}\right),
\end{equation}
with $t_{DH}$ the time at which we turn on the damped harmonic gauge and $\tau_m = 100M$.

With these gauge choices we solve a first order representation of the generalized harmonic system (Eq.~\ref{eqn:efe_gh}) in which the fundamental variables are the spacetime metric $ g_{ab}$, its spatial first derivatives $\phi_{iab}$ and its first derivatives in the direction normal to the $t=$ constant slice $\Pi_{ab} = t^c\partial_c g_{ab}$. The generalized harmonic first order system is then:

\begin{align}
  \partial_t g_{\alpha\beta} &- (1 + \gamma_1)\beta^k\partial_k g_{\alpha\beta} = -\alpha\Pi_{\alpha\beta} - \gamma_1\beta^i\Phi_{i\alpha\beta},  \\
  \partial_t\Pi_{\alpha\beta} &- \beta^k\partial_k\Pi_{\alpha\beta} + \alpha \gamma^{ki}\partial_k\Phi_{i\alpha\beta} - \gamma_1\gamma_2\beta^k\partial_k g_{\alpha\beta} \\
  &= 2\alpha g^{\zeta\delta}(\gamma^{ij}\Phi_{i\zeta\alpha}\Phi_{j\delta\beta} - \Pi_{\zeta\alpha}\Pi_{\delta\beta} -  g^{\epsilon\sigma}\Gamma_{\alpha\zeta\epsilon}\Gamma_{\beta\delta\sigma}) \notag\\
  &-2\alpha\nabla_{(\alpha}H_{\beta)} - \frac{1}{2}\alpha t^\zeta t^\delta\Pi_{\zeta\delta}\Pi_{\alpha\beta} - \alpha t^\zeta\Pi_{\zeta i}\gamma^{ij}\Phi_{j\alpha\beta} \notag\\
  &+\alpha\gamma_0[2\delta^c_{(\alpha}t_{\beta)} -  g_{\alpha\beta}t^\zeta](H_\zeta + \Gamma_\zeta) - \gamma_1\gamma_2\beta^i\Phi_{i\alpha\beta}\notag\\
    &-2\alpha(T_{\alpha\beta} - \frac{1}{2} g_{\alpha\beta}T^{\zeta\delta} g_{\zeta\delta}),\notag \\
  \partial_t\phi_{i\alpha\beta} &- \beta^k\partial_k\phi_{i\alpha\beta} + \alpha\partial_i\Pi_{\alpha\beta} - \gamma_2\alpha\partial_i g_{\alpha\beta} \\
  & = \frac{1}{2}\alpha t^\zeta t^\delta\Phi_{i\zeta\delta}\Pi_{\alpha\beta} + \alpha\gamma^{jk}t^\zeta\Phi_{ij\zeta}\Phi_{k\alpha\beta} - \alpha\gamma_2\Phi_{i\alpha\beta}\notag,
\end{align}
This amounts to a symmetric hyperbolic system of 50 coupled nonlinear equations. The constraint damping parameters $\{\gamma_0,\gamma_1,\gamma_2\}$ are additional free parameters which dampen any constraint violating modes which may grow due to small numerical errors in the evolution. To set the $\gamma$ parameters we use a trial-and-error process which concludes when we find that the constraint-violating modes do not grow significantly over time. In practice, we set the parameters to
\begin{align}
\gamma_0 &= \frac{.005}{M}+\frac{.2}{M_1}f(r^i_1, 2.5R_1)\notag\\
&+\frac{.2}{M_2}f(r^i_2, 2.5R_2)+\frac{.075}{M_2}f(r^i_c, 2.5d), \\
\gamma_1 &= 0.999 (f(r_c, 10d) - 1), \\
\gamma_2 &= \frac{.005}{M}+\frac{3}{M_1}f(r^i_1, 2.5R_1)\notag\\
&+\frac{3}{M_2}f(r^i_2, 2.5R_2)+\frac{.075}{M_2}f(r^i_c, 2.5d), \\
f(r^i&,w) = \exp(-|r^i|^2/w^2),
\end{align}
where $r^i_{1,2,c}$ correspond to the coordinate locations of the first neutron star, second neutron star and center of mass respectively, $R_1/M_1$ is the radius and mass of the 1st neutron star, $R_2/M_2$ is the radius and mass of the second neutron star, $M = M_1 + M_2$ and $d$ is the separation of the two neutron star centers.

\subsection{Neutrino Evolution}

We use a gray two moment scheme for neutrino transport introduced in \cite{foucart2016impact}. By evolving the number density of neutrinos in addition to their energy and flux densities, this new scheme guarantees exact conservation of the total lepton number. It also provides is with a local estimate of the average neutrino energy, an important quantity given the strong dependence of neutrino-matter cross-sections on the energy of neutrinos. In this section we will give an overview of this algorithm.

For each species of (anti)neutrino $\nu_i$ we can describe the neutrinos by their distribution function $f_{\nu}(x^\mu, p^\mu)$ where $x^\mu = (t, x^i)$ gives the time and the position of the neutrinos and $p^\alpha$ is the 4-momentum of the neutrinos. The distribution function $f_{(\nu)}$ evolves according to the Boltzmann equation:
\begin{equation}
  p^\alpha\left[ \frac{\partial f_{(\nu)}}{\partial x^\alpha} - \Gamma^\beta_{\alpha\gamma}p^\gamma\frac{\partial f_{(\nu)}}{\partial p^\beta}\right] = C\left[f_{(\nu)}\right],
\end{equation}
where the right-hand side includes all collisional processes (emissions, absorptions, scatterings).
In general, this is a 7-dimensional problem which is extremely expensive to solve numerically. Approximations to the Boltzmann equation have thus been developed for numerical applications. In this work, we consider the moment formalism developed by Thorne \cite{1981mnras.194..439t} and Shibata et al.~\cite{shibata:11}, in which only the lowest moments of the distribution function in momentum space are evolved.  We limit ourselves to the use of this formalism in the gray approximation, that is we only consider energy-integrated moments.  We will consider three independent neutrino species: the electron neutrino $\nu_e$, the electron antineutrinos $\bar \nu_e$, and the heavy lepton neutrinos $\nu_x$. The latter represents the 4 species $(\nu_\mu, \bar \nu_\mu, \nu_\tau, \bar \nu_\tau)$. This merging is justified because the temperatures and neutrino energies reached in our merger calculations are low enough to suppress the formation of the corresponding heavy leptons. The presence of heavy leptons would then require including the charged current neutrino interactions that differentiate between these individual species.

In the gray approximation with the first two moments of the distribution function, we evolve for each species projections of the stress-energy tensor of the neutrino radiation $T^{\mu\nu}_{rad}$. We write

\begin{equation}
T^{\mu\nu}_{rad}  = Ju^\mu u^\nu + H^\mu u^\nu + H^\nu u^\mu + S^{\mu\nu},
\end{equation}
with $H^\mu u_\mu = S^{\mu\nu}u_\mu = 0$ and $u^\mu$ the 4-velocity of the fluid. We can decompose the momentum as follows

\begin{equation}
  p^\alpha = \nu(u^\alpha + l^\alpha),
\end{equation}

with $l^\alpha u_\alpha = 0$ and $l^\alpha l_\alpha = 1$. With this decomposition, the energy $J$, flux $H^\mu$ and stress tensor $S^{\mu\nu}$ of the neutrino radiation as observed by an observer comoving with the fluid are related to the neutrino distribution function by

\begin{align}
  J &= \int^\infty_0 \mathrm{d}\nu \nu^3 \int \mathrm{d}\Omega f_{(\nu)}(x^{\alpha}, \nu, \Omega) \\
  H^\mu &= \int^\infty_0 \mathrm{d}\nu \nu^3 \int \mathrm{d}\Omega f_{(\nu)}(x^{\alpha}, \nu, \Omega) l^{\mu}\\
  S^{\mu\nu} &= \int^\infty_0 \mathrm{d}\nu \nu^3 \int \mathrm{d}\Omega f_{(\nu)}(x^{\alpha}, \nu, \Omega) l^{\mu}l^\nu,
\end{align}
where $\nu$ is the neutrino energy in the fluid frame, and $\int\mathrm{d}\Omega$ denotes integrals over solid angle on a unit sphere in momentum space.
We also utilize the decomposition of $T^{\mu\nu}_{rad}$ in terms of the energy, flux and stress tensor observed by an inertial observer (i.e. an observer whose worldline is tangent to the normal vector $n^\mu$),

\begin{equation}
  T^{\mu\nu}_{rad} = En^\mu n^\nu + F^\mu n^\nu + F^\nu n^\mu + P^{\mu\nu},
\end{equation}
with $F^\mu n_\nu = P^{\mu\nu}n_\mu = F^t = P^{t\nu} = 0$.%, and $n^\alpha$ the unit normal to a $t$ = constant slice.

We define a projection operator on the reference frame of an observer comoving with the fluid:
\begin{equation}
  h_{\alpha\beta} = g_{\alpha\beta} + u_\alpha u_\beta.
\end{equation}
We can then write equations relating the fluid frame moments to the inertial frame moments

\begin{align}
  E &= W^2 J + 2 W v_\mu H^\mu + v_\mu v_\nu S^{\mu\nu}, \\
  F_\mu &= W^2 v_\mu J + W(g_{\mu\nu} - n_\mu v_\nu) H^\nu \\
  &+ W v_\mu v_\nu H^\nu + (g_{\mu\nu} - n_\mu v_\nu) H^\nu + Wv_\mu v_\nu H^\nu \notag\\
  &+ (g_{\mu\nu} + n_\mu v_\nu) v_\rho S^{\nu\rho}, \notag\\
  P_{\mu\nu} &= W^2 v_\mu v_\nu J + W(g_{\mu\nu} - n_\mu v_\rho)v_\nu H^\rho \\
  &+ (g_{\mu\rho} - n_\mu v_\rho)(g_{\nu\kappa} - n_\nu v_\kappa)S^{\rho\kappa} \notag \\
  &+ W(g_{\rho\nu} - n_\rho v_\nu) v_\mu H^\rho, \notag
\end{align}
using the decomposition of the 4-velocity
\begin{equation}
  \label{eqn:vel_decomp}
  u^\mu = W(n^\mu + v^\mu),
\end{equation}
where $v^\mu n_\mu = 0$ and $W = \sqrt{1 + v_i v^i}$.

By taking moments of the Boltzmann equation, the evolution equations for $\tilde E = \sqrt{\gamma}E$ and $\tilde F^i = \sqrt{\gamma} F^i$ can then be written in conservative form

\begin{align}
  \partial_t \tilde E &+ \partial_j (\alpha \tilde F^j - \beta^j \tilde E) \\
  &= \alpha(\tilde P^{ij} K_{ij} - \tilde F^j \partial_j \ln \alpha - \tilde S_\text{rad}^\alpha n_\alpha), \notag\\
  \partial_t \tilde F_i &+ \partial_j (\alpha \tilde P_i^j - \beta^j \tilde F_i) \\
  &= (-\tilde{E} \partial_i \alpha + \tilde{F}_k \partial_i \beta^k + \frac{\alpha}{2} \tilde{P}^{jk} \partial_i \gamma_{jk} + \alpha \tilde{S}_{\text{rad}}^\alpha \gamma_{i\alpha}) \notag,
\end{align}
where $\gamma$ is the determinant of $\gamma_{ij}$, $\tilde{P}_{ij} = \sqrt{\gamma}P_{ij}$, and $\tilde{S}_{rad}^\alpha = \sqrt{\gamma} S_{rad}^\alpha$ includes all collisional source terms. Additionally, we consider the number current density for each species of neutrino:

\begin{equation}
  N^\mu = Nn^\mu + \mathcal{F}^\mu,
\end{equation}
where N is the number density of neutrinos, and $\mathcal{F}^\mu$ the number flux density. We can define the number current in the fluid frame by

\begin{equation}
  N^\mu = \frac{Ju^\mu}{\langle \nu \rangle} + \frac{H^\mu}{\langle \nu^F \rangle},
  \label{eq:number_current}
\end{equation}
where $\langle \nu \rangle$ is the average neutrino energy and $\langle \nu^F \rangle$ is the flux-weighted average neutrino energy. Using Eq.~\ref{eq:number_current} we can get an estimate for the energy

\begin{equation}
\langle \nu \rangle = W\frac{(E-F_iv^i)}{N},
\end{equation}
where $W$ is the Lorentz factor introduced in Eq.~\ref{eqn:vel_decomp}. The evolution equation for $\tilde{N} := \sqrt{\gamma}N$ is

\begin{equation}
  \partial_t \tilde N + \partial_j(\alpha \sqrt{\gamma} \mathcal{F}^j - \beta^j \tilde N) = \alpha \sqrt{\gamma} C_{(0)},
\end{equation}
where the source term accounts for neutrino-matter interactions.

To close this system of equations we need three additional ingredients: a prescription for the computation of $P^{ij}(E, F_i)$ which is called the closure relation, a prescription for the computation of the number flux $\mathcal{F}^j$ (specific to te evolution of the number density N in this paper) and the collisional source terms $\tilde{S}^\alpha_{\rm rad}, C_{(0)}$.

For $P^{ij}(E, F_i)$  we interpolate between optically thick and thin limits:
\begin{equation}
P^{ij} = \frac{3p-1}{2}P^{ij}_{\text{thin}} + \frac{3(1-p)}{2}P^{ij}_{\text{thick}}.
\end{equation}

Here the parameter p is known as the variable Eddington factor and our choice for the functional form of p in terms of the lower moments $H,J$ is known as the Minerbo closure~\cite{minerbo1978}. Our choices for $P^{ij}_{\rm thin}, P^{ij}_{\rm thick}$ and $p$ are discussed in the Appendix of \cite{foucartm1:2015}.

For $\mathcal{F}^i$, by definition we have

\begin{equation}
  \mathcal{F}^i = \frac{JWv^i}{\langle \nu \rangle} + \frac{\gamma^i_\mu H^\mu}{\langle \nu^F \rangle}
\end{equation}
where the flux-weighted average neutrino energy,  $\langle \nu^F \rangle$ is computed in such a way to take the effects of a finite optical depth on the spectrum into account, see \cite{foucart:2015gaa} for details.

For the source terms $\tilde{S}^\alpha$, we assume that the fluid has an energy-integrated emissitivity $\bar \eta$ due to the charged-current reactions
\begin{align}
  p + e^- &\rightarrow n + \nu_e, \\
  n + e^+ &\rightarrow p + \bar{\nu}_e,
\end{align}
as well as electron-positron pair annihilation
\begin{equation}
  e^+ + e^- \rightarrow \nu_i \bar \nu_i,
\end{equation}
plasmon decay
\begin{equation}
  \gamma \rightarrow \nu_i\bar \nu_i,
\end{equation}
and nucleon-nucleon Bremsstrahlung
\begin{equation}
  N + N \rightarrow N + N + \nu_i + \bar \nu_i.
\end{equation}
The inverse reactions are responsible for an energy-averaged absorption opacity $\bar \kappa_a$. We also consider an energy-averaged scattering opacity $\bar \kappa_s$ due to elastic scattering of neutrinos on nucleons and heavy nuclei. Neglecting other reactions (e.g. inelastic scatterings and $\nu\bar\nu$ annihilation) the source terms $S^\alpha$ can then be shown to be \cite{shibata2011truncated}
\begin{equation}
  \tilde S_\text{rad}^\alpha = \sqrt{\gamma} [\bar \eta u^\alpha - \bar \kappa_\alpha J u^\alpha - (\bar \kappa_a + \bar \kappa_s) H^\alpha].
\end{equation}
The collisional source term for the number density $\bar N$ is given by
\begin{equation}
  C_{(0)} = \bar \eta_N - \bar \kappa_N \frac{J}{\langle\nu\rangle} = \bar \eta_N - \frac{\bar \kappa_N J \bar N}{W(\bar E - \bar F_iv^i)}.
\end{equation}
Thus we need the energy integrated emissivities $(\bar \eta, \bar \eta_N)$ and the energy-averaged opacities $(\bar \kappa_A, \bar \kappa_S, \bar \kappa_N)$ to compute the source terms. 
We begin by computing the energy-integrated energy and number emissivities $(\bar \eta, \bar \eta_N)$ and energy-averaged equilibrium opacities $(\bar \kappa^{eq}_A, \bar \kappa^{eq}_S)$, which assume a thermal distribution of neutrinos in equilibrium with the fluid. The emissivities are computed following Ruffert et al.~\cite{ruffert1996}, except for nucleon-nucleon Bremsstrahlung for which the emissivity is computed following Burrows et al.~\cite{burrows2006b}. The equilibrium opacities are computed using Kirchoff's Law:

\begin{equation}
\bar \eta = \bar\kappa^{eq}\int B_{(\nu)}\mathrm{d}\nu,
\end{equation}

where $B_{(\nu)}$ is the blackbody spectrum in equilibrium with the fluid. This ensures that, in regions where neutrinos are in thermal equilibrium with the fluid, they have the correct energy density $\bar \eta/\bar\kappa^{eq}$. To account for the fact that the average energy of neutrinos may be very different from its equilibrium value, we then correct the opacities by making use of the fact that processes computed here have cross-sections scaling as $T_\nu^2$:

\begin{align}
   \bar \kappa_A &= \bar \kappa^{eq}_A\frac{T_\nu^2}{T_{\text{fluid}}^2}, \\
  \bar \kappa_S &= \bar \kappa^{eq}_S\frac{T_\nu^2}{T_{\text{fluid}}^2},
\end{align}
where $T_\nu$ is the neutrino temperature computed from the neutrino energy and number density, assuming a blackbody spectrum (see \cite{foucart2016impact} for more details). Lastly we set
\begin{equation}
\bar \kappa_N = \bar \kappa_A \frac{\bar\eta_N}{\bar\eta}\frac{F_3(\eta_\nu)T_{\rm fluid}}{F_2(\eta_\nu)},
\end{equation}
where $\eta_\nu = \mu_\nu/T$, $\mu_\nu$ is the chemical potential of neutrinos in equilibrium with the fluid, and $F_k(\eta_\nu)$ is the Fermi integral
\begin{equation}
F_k(\eta_\nu) = \int^\infty_0\frac{x^k}{1 + \exp(x - \eta_\nu)}\mathrm{d}x.
\end{equation}
This choice ensures that the average energy of neutrinos in equilibrium with the fluid takes its expected value,
\begin{equation}
<\nu>^{eq} = \frac{\bar \eta}{\bar \kappa_A^{\rm eq}} \frac{\bar \kappa_N^{\rm eq}}{\bar \eta_N} = \frac{F_3(\eta_\nu)}{F_2(\eta_\nu)} T_{\rm fluid}.
\end{equation}

\subsection{Fluid Evolution}

The neutron stars are described by ideal fluids with stress tensor
\begin{equation}
T_{\mu\nu} = \rho_0 h u_\mu u_\nu + Pg_{\mu\nu},
\end{equation}
where $\rho_0$ is the rest mass density, h the specific enthalpy, P the pressure and $u^\mu$ the 4-velocity. The general relativistic equations of hydrodynamics are evolved in conservative form, using the conservative variables
\begin{align}
  %% \rho_* &= \rho_0W\sqrt{\gamma}, \\ %-\sqrt{\gamma}n_\mu u^mu /rho_0 \\
  \rho_* &= -\sqrt{\gamma}n_\mu u^\mu \rho_0, \\
  %% \tau &= \rho_* (hW - 1) - \sqrt{\gamma}P, \\ %\sqrt{\gamma}n_\mu n_\nu T^{\mu\nu} - \rho_* \\
  \tau &= \sqrt{\gamma}n_\mu n_\nu T^{\mu\nu} - \rho_*, \\
  %% S_k &= \rho_* h u_i, %-\sqrt{\gamma}n_\mu T^\mu_k
  S_k &= -\sqrt{\gamma}n_\mu T^\mu_k.
\end{align}
The 3 + 1 stress-energy conservation equations for radiation-hydrodynamics $\nabla_\nu T^{\mu\nu} = -S^\mu_\text{rad}$ and the baryon conservation equations $\nabla_\mu (\rho_0 u^\mu) = 0$ then become \cite{shibata2011truncated}
\begin{align}
  \partial_t \rho_* + \partial_j(\rho_* v_t^j) =&\:0, \\
  \partial_t \tau + \partial_i (\alpha^2 \sqrt{\gamma} T^{0i} - \rho_* v_t^i) =& -\alpha \sqrt{\gamma} T^{\mu\nu} \nabla_\nu n_\mu\\
  &+ \alpha \tilde S_\text{rad}^\alpha n_\alpha, \notag\\
  \partial_t S_i + \partial_j(\alpha\sqrt{\gamma}T^j_i) =&\: \frac{1}{2}\alpha \sqrt{\gamma} T^{\mu\nu} \partial_i \gamma_{\mu\nu} \\
  &- \alpha \tilde S_\text{rad}^\alpha \gamma_{i\alpha},\notag
\end{align}

where $v_t^i = u^i/u^0$ is the transport velocity.

We evolve these equations using standard high-order shock-capturing finite volume method. We compute values of the physical quantities $(\rho_0,T,u_i)$ on cell faces from their cell-average values using fifth order WENO5 reconstruction~\cite{borges}, and compute numerical fluxes on faces using the approximate HLL Riemann solver~\cite{hll}.

\subsection{Composition Evolution}

The fluid composition in our simulations is described by the electron fraction:

\begin{equation}
  Y_e = \frac{n_p}{n_p + n_n},
\end{equation}
where $n_p$ and $n_n$ are the proton and neutron number densities, respectively (the net electron number density $n_{e^{-}} - n_{e^{+}} = n_p$, due to charge neutrality in the fluid). From lepton number conservation, we have:

\begin{equation}
  \partial_t \left(\rho_* Y_e \right) + \partial_i\left(\rho_*Y_e v_t^i\right)= -\sum_{\nu}\textrm{sign}(\nu)\alpha \sqrt{\gamma} C^{\nu}_{(0)},
\end{equation}
where $\sum_{\nu}$ sums over neutrino species with $\textrm{sign}(\nu)$ set to 1 for $\nu_e$, -1 for $\bar \nu_e$, and 0 for $\nu_x$. Importantly for the electron fraction, evolving the neutrino number density $N$ frees us from having to guess at the average neutrino energy when computing the coupling to the fluid. It also guarantees that the source term for the evolution of the electron fraction of the fluid is fully consistent with the evolution of the neutrino number density, thus conserving the total lepton number of the system \cite{foucart2016impact}. 

\subsection{Equation of State (EOS)}

We use three finite-temperature, composition dependent nuclear-theory based equations of state. Two of them are based on relativistic mean field (RMF) models  \cite{walecka1974theory} and one based on the single nucleus approximation for heavy nuclei \cite{lattimer1991generalized}. They are:

\begin{enumerate}
\item{DD2 \cite{hempel:2011mk}: This EOS is based on nuclear statistical equilibrium with a finite volume correction coupled to a relativistic mean field theory for treating high-density nuclear matter. DD2 contains neutrons, protons, light nuclei such as deuterons, helions, tritons and alpha particles and heavy nuclei. DD2 does not satisfy the so-called flow-constraint \cite{hempel2017well}.}
\item{
LS220 \cite{lattimer1991generalized}: is based on the single nucleus approximation
for heavy nuclei where the thermal distribution of different nuclear species is replaced by a single representative heavy nucleus. LS220 contains  neutrons, protons, alpha particles and heavy nuclei. LS220 does not satisfy the constraints from Chiral effective field theory \cite{hempel2017well}.
}
\item{SFHo \cite{2013apj...765l...5s}: This EOS, like DD2, also uses an RMF model, containing neutrons, protons, light nuclei such as deuterons, helions, tritons and alpha particles and heavy nuclei. However, SFHo uses a different RMF parameterization,
specifically designed to match neutron star properties as inferred by observations. SFHo shows some minor
deviations from Chiral effective field theory calculations \cite{hempel2017well}.
}
\end{enumerate}

\begin{figure}
  \includegraphics[width=.45\textwidth]{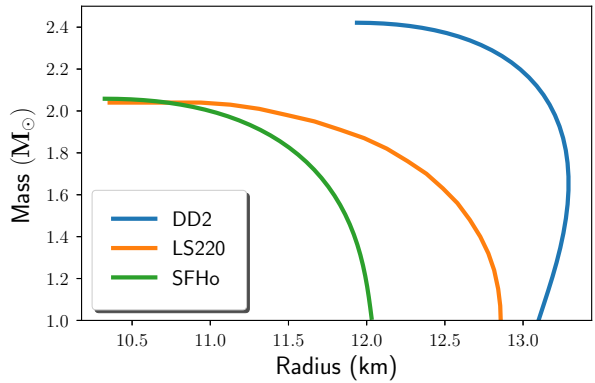}
\caption{
  M-R curves for each of the equations of states used in this work.
}
\label{fig:eos_mr}
\end{figure}

While each EOS shows deviations from theoretical calculations, they all have a radius, maximum mass and tidal deformalibity that are compatible with observations of isolated neutron stars~\cite{demorest:2010bx,hempel2017well}. The LS220 and SFHo EOSs are also compatible with all constraints derived from the gravitational wave signal emitted by GW170817~\cite{gw170817-pe,gw170817-nsradius}, while the DD2 equation of state has a tidal deformability for $1.4M_\odot$ neutron stars just outside of the strictest bound derived using parametrized equation of state models~\cite{gw170817-nsradius}. For all three equations of state, we show mass-radius curves in Figure~\ref{fig:eos_mr}. These curves were computed by integrating the TOV equations for the EOS by using the table values at $T=0.1$\,MeV in beta equilibrium. From Figure~\ref{fig:eos_mr}, we can clearly see that for each EOS there is a maximum mass for non-rotating isolated neutron stars. Furthermore we can associate with each EOS an average density at this maximum mass as $\langle \rho \rangle := 3M_{max}/4\pi R^3_{max}$ and the ratio of this average density and the central density $\rho_c$ of the star is an indication of how ``stiff'' the EOS is. Higher values of $\langle \rho \rangle/\rho_c$ correspond to stiffer EOS, while lower values correspond to softer EOS \cite{bauswein2013prompt}. In our case, from stiffest to softest, the EOS are ranked DD2, LS220 and SFHo. Softer EOS tend to make smaller stars at a fixed mass. For a fiducial mass of $1.4M_\odot$, the radii are 13.2 km, 12.7 km, 11.9 km for DD2, LS220 and SFHo respectively. We can also introduce a useful quantity called the compactness, which is defined by $C=M/R$. Softer EOS tend to make more compact stars. The maximum mass however is not a good indicator of whether the post-merger remnant will promptly collapse to a black hole (BH) after merger because the remnant will most likely be differentially rotating and held up by centrifugal and thermal forces against collapse.
A more accurate estimate for the maximum mass at which the merger promptly collapses to a BH is given by Bauswein~{\it et al.}~\cite{bauswein2013prompt} who ran simulations of BNS mergers with a smoothed particle hydrodynamics code that employs the conformal flatness approximation of the Einstein field equations and includes a GW backreaction scheme to determine a thershold mass for prompt collapse across 12 different realistic EOS. Bauswein et al. found the threshold masses for SFHo, LS220 and DD2 were $2.95M_\odot$, $3.05M_\odot$ and $3.35M_\odot$ respectively. Thus the SFHo and LS220 models are more likely to undergo prompt collapse for the higher end of the total masses considered in this paper ($\sim 2.9M_\odot$, see Sec.~\ref{sec:initial_models}).

\subsection{Initial Models}
\label{sec:initial_models}
We extend our previous work \cite{foucart:2015gaa} and study the merger
of unequal mass neutron star binaries with the neutrino transport scheme introduced in \cite{foucart2016impact} and discussed above. We study mass ratios between .76 and 1 and masses between 1.2$M_\odot$ and 1.56$M_\odot$ which are both within the ranges of current binary neutron star observations (see e.g. \cite{lattimer:2012nd}). We focus here on the late stages of the coalescence, comprising the last $3-5$ orbits (depending on EoS) before merger and we stop evolving any systems once they collapse. The parameters for the initial models and the type of post-merger remnants they create (before or as of 7.5-ms post-merger) are shown in Table \ref{tab:initial_models}. The specific grid setup for each model will be discussed in the following section.

\begin{table}
\centering
\begin{tabular}{cccccccccl} \toprule
Model & EOS & \(M_{1}\) & \(M_{2}\) & \(q\) & $C_1$ & $C_2$ & $d_0 (km)$  & Collapse?\\ \midrule
D144144 & DD2 & 1.44 & 1.44 & 1.0 & .161 & .161 & 48.7 & No \\
D12132 & DD2 & 1.2 & 1.32 & .91 & .134 & .147 & 48.7 & No\\
D12144 & DD2 & 1.2 & 1.44 & .83 & .134 & .161 & 48.7 & No\\
D12156 & DD2 & 1.2 & 1.56 & .77 & .134 & .173 & 48.7 & No\\
 \midrule
L144144 & LS220 & 1.44 & 1.44 & 1.0 &.175  &.175 & 44.3 & $(\sim 2 ms)$\\
L12132 & LS220 & 1.2 & 1.32 & .91 & .146 & .161  & 44.3 & No \\
L12144 & LS220 & 1.2 & 1.44 & .83 & .146  &.175  & 44.1 & No\\
L12156 & LS220 & 1.2 & 1.56 & .77 & .146  & .191 & 44.3 & $(\sim 4.5 ms)$\\
 \midrule
S144144 & SFHo & 1.44 & 1.44 & 1.0 &.179  & .179 & 44.3 & $(\sim .5 ms)$\\
S12132 & SFHo & 1.2 & 1.32 & .91 & .148  & .163  & 44.3 & No\\
S12144 & SFHo & 1.2 & 1.44 & .83 & .148  & .179 & 44.3 & No\\
S12156 & SFHo & 1.2 & 1.56 & .77 & .148  & .195 & 44.3 & $(\sim 1.4 ms)$\\
 \bottomrule
\end{tabular}
\caption{Parameters for the initial models presented in this paper. $M_1$ and $M_2$ are the gravitational masses of the two neutron stars, $C_1$ and $C_2$ are the compactness of the respective neutron stars and $d_0$ is the initial coordinate distance between the centers of stars. The last column gives the post-merger remnant as of 7.5-ms post-merger. If the binary collapsed before 7.5-ms a rough estimate for the time of collapse is shown in parenthesis. We have codenamed each simulation by joining the first letter of the EOS name with the digits of the gravitational masses for that system, e.g. the DD2 $1.2M_\odot + 1.44M_\odot$ run is abbreviated D12144 throughout the paper.}
\label{tab:initial_models}
\end{table}

\subsection{Grid Setup}

Before the two neutron stars enter into contact, the pseudospectral grid on which we evolve Einstein's equations takes advantage of the approximate spherical symmetry of the neighborhood of each star, and in the far-field region. The evolved spatial slice is decomposed into two small balls around the center of each neutron star, sets of spherical shells around each star, spherical shells in the wave-zone region far from the stars and distorted cubes to connect the three spherically symmetric regions. The inner ball is expanded into Zernike polynomials, the shells into Chebyshev polynomials (in radius) and spherical harmonics (in angle) and the distorted cubes in Chebyshev polynomials. The grid follows the centers of the neutron stars defined as the center of mass of the matter in the $x < 0$ and $x > 0$ half planes, through a simple rotation and scaling of the grid coordinates. %We do not exploit the reflection symmetry $z \rightarrow -z$ in our runs.

We maintain this grid decomposition for the evolution of Einstein's equations up to the point at which the maximum density on the grid increases beyond the low-level oscillations observed during the inspiral. This rise in the density signifies the transition from two well-separated neutron star cores to a single, more massive object. At that point we switch to a grid which is fully centered on the coordinate center of mass of the system. This grid  is made of a ball at the origin of the coordinate system, surrounded by 59 spherical shells extending to the outer edge of the computational domain. Both before and after merger, the outer boundary is located at $40d_0$, with $d_0$ the initial separation of the binary, provided in Table~\ref{tab:initial_models}.

The finite volume grid on which we evolve the general relativistic equations of hydrodynamics is very simple. Before the two neutron stars get into contact, it is composed of two cubes, each centered on a neutron star and composed of $96^3$ cells.  In the coordinate system comoving with the neutron star centers, the neutron stars expand as the binary inspirals. To avoid losing matter to the outer boundary of the finite volume grid, we expand the grid by $4.5\%$ every time the flux of matter across the outer boundary exceeds $0.015 M_\odot s^{-1}$. As the inspiral lasts less than 10 ms, this implies a mass loss well below $10^{-4} M_\odot$ before merger. As the two neutron stars approach each other, the two finite volume boxes will eventually intersect. During merger, we would like to follow the forming massive neutron star remnant, the tidal tails, the accretion disk, and any ejected material. We switch to a finite difference grid centered on the forming remnant with 3 levels of refinement. Each level has, at our standard resolution $200^2 \times 100$ cells, with the finest grid spacing listed in Table~\ref{tab:grid_setup} and each coarser level increasing the grid spacing by a factor of 2. The lower number of cells in the vertical direction reflects the fact that the remnant is less extended in that direction, and thus that we do not need the finest grid to extend as far vertically as horizontally.

\begin{table}
\centering
\begin{tabular}{ccc}\toprule
Name & $\mathrm{d}x_{\text{ins}}$ (m) & $\mathrm{d}x_{\text{mer}}$ (m) \\ \midrule
D12132 & 279 & 300 \\
D12144 & 280 & 300 \\
D12156 & 279 & 300 \\
D144144 & 273 & 300 \\ \midrule
S12132 & 251 & 300 \\
S12144 & 250 & 300 \\
S12156 & 238 & 300 \\
S144144 & 238 & 300 \\ \midrule
L12132 & 252 & 300 \\
L12144 & 253 & 300 \\
L12156 & 252 & 300 \\
L144144 & 245 & 300 \\ \bottomrule
\end{tabular}
\caption{Finite difference grid sizes for the initial models. All simulations were set so that they would have a 300m resolution during merger. For reference, the DD2 $1.44M_\odot$ star has a $\sim 10.48 km$ radius in our grid coordinates.
}
\label{tab:grid_setup}
\end{table}

\subsection{Ejecta Analysis}

In a BNS merger, matter expelled at high velocity may
ultimately become unbound from the central gravitational potential. Two indicators have been used to label matter unbound:
\begin{enumerate}
\item{$\mathbf{u_t < -1}$: For a stationary spacetime, $u_t$ (the projection of the 4-velocity along the timelike Killing vector field) is a constant of motion for geodesics. Assuming the space is also asymptotically flat, the Lorentz factor $W$ satisfies $W = -u_t$ at infinity. Therefore we may flag matter as unbound using the condition  $u_t < -1$. This assumes that the outflow is made of isolated particles following geodesics, thus neglecting the impact of pressure gradients and r-process heating on the fluid. It also assumes that the metric is time-independent. Neither assumption is entirely correct in the dynamical spacetime of a post-merger remnant.} \\
\item{$\mathbf{hu_t < -1}$: For a stationary relativistic fluid flow, the relativistic Bernoulli equation \cite{rezzolla2013relativistic} implies $hu_t$ is constant along fluid worldlines. In an asymptotically flat spacetime, we would expect $W = -u_t$ (if the fluid particles follow geodesics). Since the relativistic enthalpy $h$ is only defined up to a constant factor which can be set such that $h \leftarrow 1$ at spatial infinity. Therefore, we may flag matter as unbound using the condition $hu_t < -1$. The main difference with the previous criteria is that all of the thermal energy of the fluid is now assumed to be transformed into kinetic energy as the fluid decompresses. Energy deposition due to r-process heating is also treated differently. We implicitly assume that the difference in binding energy between particles in nuclear statistical equilibrium (NSE) at the current density, temperature, and composition, and their binding energy at low density and temperature but for the same $Y_e$, is entirely deposited/removed from the fluid's kinetic energy. This neglects significant out-of-NSE evolution during r-process nucleosynthesis (see~\cite{foucartbhns2016})}.
\end{enumerate}

We have found previously \cite{foucart:2015gaa} that the second indicator $hu_t < -1$, the Bernoulli criterion, produces
qualitatively more accurate results in SpEC, therefore all material labelled unbound in this paper uses this indicator. It is important to note that no indicator is exact.  In fact, the Bernoulli criterion has been shown to result in as much as twice the ejected matter as the $u_t$ condition~\cite{kastaun:2014fna}. 

With the Bernoulli criterion, we flag matter that is still on the grid if it is at least $50M_\odot$ away from the center of the remnant and compute the ejected mass as follows
\begin{equation}
M^{on}_{ej}(t) = \int_{r>50M_\odot}\rho_0 W\mathcal{H}(-hu_t - 1)\sqrt{\gamma}{d}^3x,
\end{equation}
where $\mathcal{H}(\cdot)$ is the Heaviside function. While the $50M_\odot$ threshold is arbitrary, we have found that it comfortably excludes any ejecta very close to the remnant which may not ultimately become unbound.

We also measure unbound material that leaves the computational grid (for each run, the grid is roughly $200M_\odot \times 200M_\odot \times 100M_\odot$) using

\begin{equation}
M^{\rm off}_{\rm ej}(t) = \int^t_0\int_S\rho_0\mathcal{H}(-hu_t - 1) W(\alpha v^i - \beta^i)n_i\mathrm{d}S\mathrm{d}t'.
\end{equation}
Here $v^i$ is the fluid 3-velocity, $W$ is the Lorentz factor, defined by $W := (1-v^iv_i)^{-1/2}$, $S$ is the grid boundary and $\rho_0 \mathcal{H}(-hu_t - 1)W(\alpha v^i - \beta^i)n_i$ represents the flux of unbound material leaving the boundary. At a time $t$, we estimate the full ejecta as

\begin{equation}
M_{\rm ej}(t) = M^{\rm on}_{\rm ej} + M^{\rm off}_{\rm ej}
\end{equation}

Finally, to compute average quantities for the ejected matter, we use the following definition of the mass-weighted average for a quantity X.

\begin{equation}
  \left< X \right> = \frac{1}{M_{\rm ej}}\int X \mathrm{d}M_{\rm ej},
\end{equation}

\subsection{Errors}

Since we performed simulations only at one resolution, it is difficult to derive error estimates. However, Hotokezaka et al. \cite{hotokezaka:13} find a $\sim 10\%$ relative error for their mass ejecta properties for unequal mass binaries and we have independently confirmed their results with our code using the same piecewise polytropes EOS they use at similar resolution. However in this work, we use tabulated EOS and have a slightly coarser resolution. Furthermore, in equal mass runs (see \cite{foucart:2015gaa}) we have found that relative errors can be as high as $\sim 50\%$. Hotokezaka et al.~\cite{hotokezaka:13} find similar variations in the mass ejected by equal mass binaries, with some systems showing little sign of convergence even at higher resolution. We therefore conclude that our errors in the ejected mass may be as high as $\sim 50\%$ and conservatively set our error as
\begin{equation}
\label{eqn:error}
  \Delta M_{ej} = .5M_{ej} + 10^{-4}M_\odot,
\end{equation}
with the lower bound $10^{-4}M_\odot$ coming from the fact that we ignore outflows of this size during the regridding in the inspiral stage. Practically speaking, this is likely to be a significant overestimate of the error for the average simulation in our dataset and is more representative of the error for the worst simulations presented here. %The main objective of this study, however, is to determine trends in the ejecta properties across EOS and mass ratio and hence Eq.~\ref{eqn:error} is not a major limitation. 
Our numerical setup also ignores magnetic fields. Over the short time scales considered here, magnetohydrodynamics (MHD) effects are not expected to affect the evolution of neutron star remnant, but could drive additional outflows from the disk \cite{kiuchi2014,neilsen2014magnetized}. Over longer time scales, magnetic fields would be critical to the spin evolution of the remnant neutron star, angular momentum transport, heating in the disk, and possibly the formation of relativistic jets and magnetically-driven outflows. General relativistic MHD simulations of postmerger disks show
that up to $\sim 40\%$ of the accretion disk 
%(around $0.013M_\odot$)  \Francois{The fractional value is more important than the IC-dependent mass}
can be ejected over 9 seconds \cite{fernandez2019long}. We 
discuss the difference magnetic fields and longterm neutrino-winds could make on ejecta estimates in Section~\ref{sec:long_term}.

\section{Numerical Results}

\subsection{General Overview}

\begin{figure*}[ht!]
  \includegraphics[width=.45\textwidth,frame]{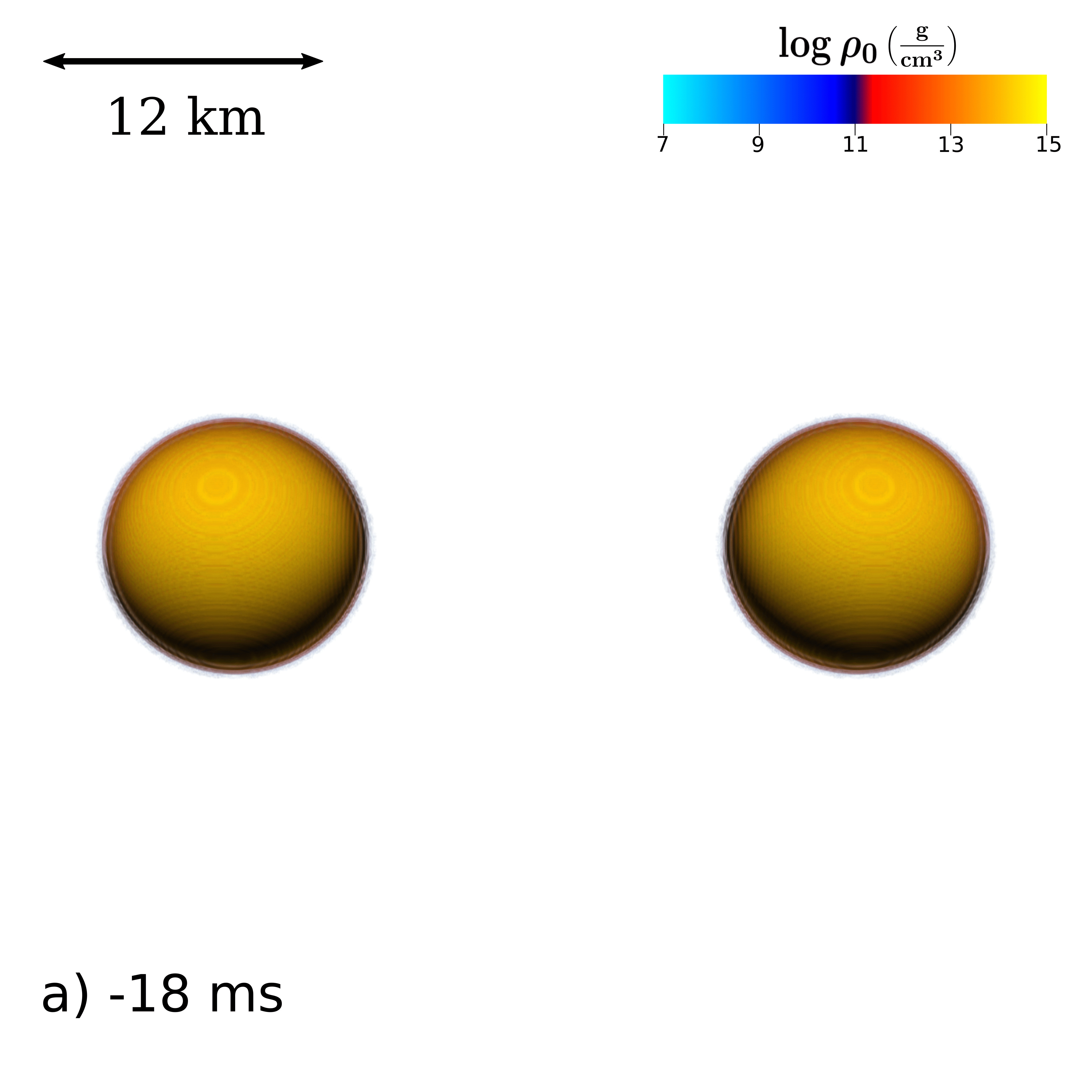}\hfill
            \includegraphics[width=.45\textwidth,frame]{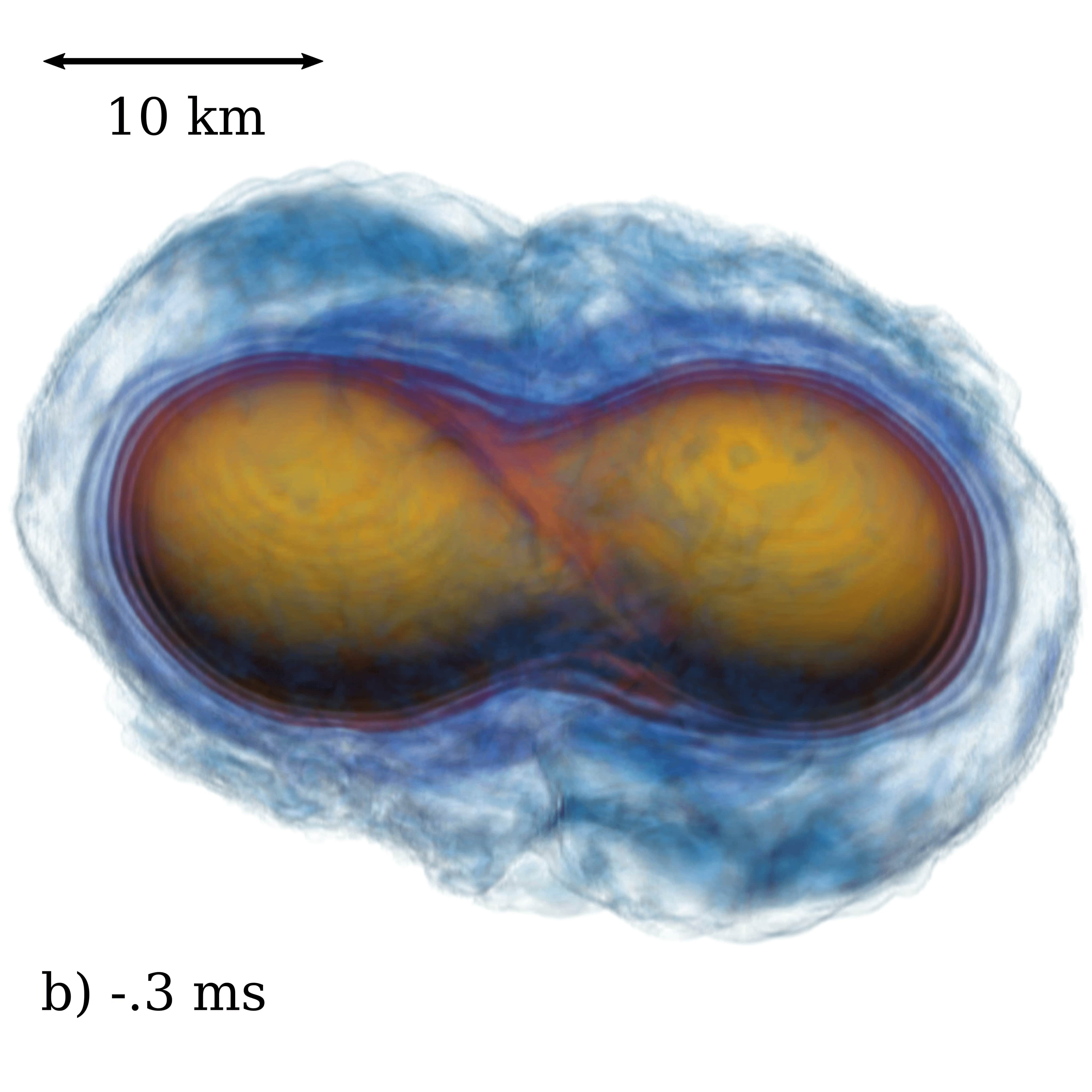}\\
            \includegraphics[width=.45\textwidth,frame]{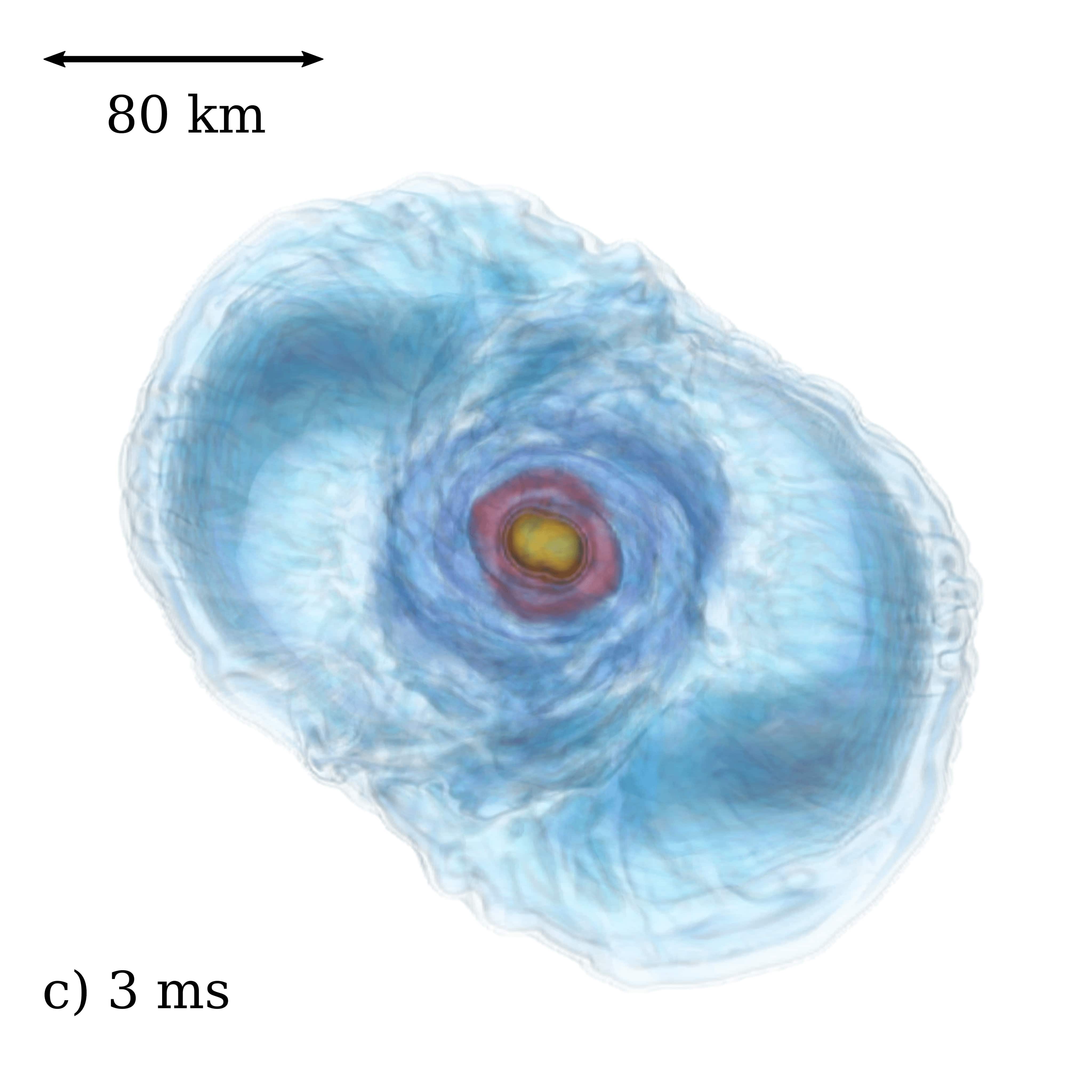}\hfill
            \includegraphics[width=.45\textwidth,frame]{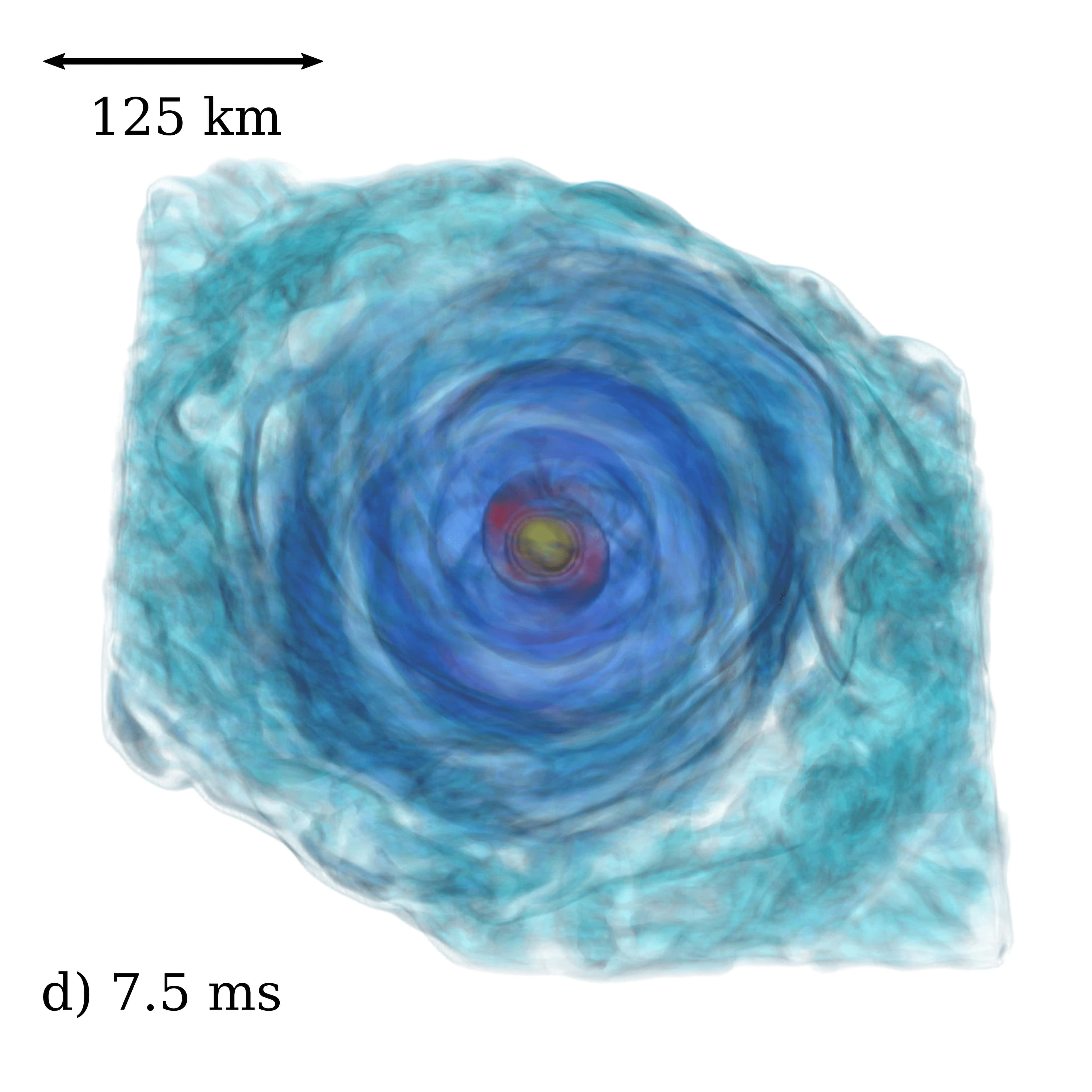}
            \caption{The evolution of rest mass density $\rho_0$ for the D144144 model. For the first $\sim$18ms the stars orbit each other before merging at $\sim$ 0ms. At around $\sim 3$ ms we can see a post-merger remnant almost fully formed with two tidal tails. At $\sim$7.5 ms the post-merger remnant has largely settled and most of the unbound material has left the grid.  }
            \label{fig:3d_panel}
        \end{figure*}

Prior to merger, which is defined as the peak of the gravitational-wave amplitude, the compact objects inspiral
around each other for 3-5 orbits, with the actual number varying between EOS. This stage is shown in first panel of Fig.~\ref{fig:3d_panel} for the DD2 $1.44M_\odot + 1.44M_\odot$ model and takes approximately eighteen milliseconds. After the stars start merging, which is shown in the second panel of Fig.~\ref{fig:3d_panel}, the properties of the system are then largely determined by the compactness and the mass-ratio of the pre-merger neutron stars. These parameters control two important mechanisms. The first mechanism is tidal forces, which rip off matter as the binary gets closer, ejecting matter at angles close to the orbital plane. The second mechanism is the contraction and recoil of the cores (called core-bounce), which produces shock waves that eject matter quasi-spherically.  From these two mechanisms, we can make predictions on expected trends. For example, as the mass-ratio increases, the less massive star gets more tidally disrupted and thus on merger, the effect of the shock on ejection of matter is lessened, but the effect of tidal torque propelling ejecta off the grid is increased. Similarly, as the EOS softens, the compactness of the star increases and the effect of core-bounces on the ejection of matter increases. In general we might expect that changing the mass-ratio might have a small effect on the emission of SFHo and LS220, but a larger effect on the ejecta of DD2 (see e.g \cite{sekiguchi2016dynamical}). As we see in the third panel of Fig~\ref{fig:3d_panel}, the DD2 EOS showcases well defined tidal tails larger than eighty kilometers in length around three milliseconds post-merger. This is to be expected from a very stiff EOS, which creates large stars that are torn apart by tidal forces in the late inspiral. A few milliseconds after this, at around 7.5 milliseconds post-merger (as shown in the fourth panel of Fig.~\ref{fig:3d_panel}), the tidal tails have left the grid and we are left with a large, circularized, 125 km radii disk surrounding the remnant with matter outflows continuously pouring off and leaving the grid.

\begin{figure*}[!htbp] \includegraphics[width=\textwidth]{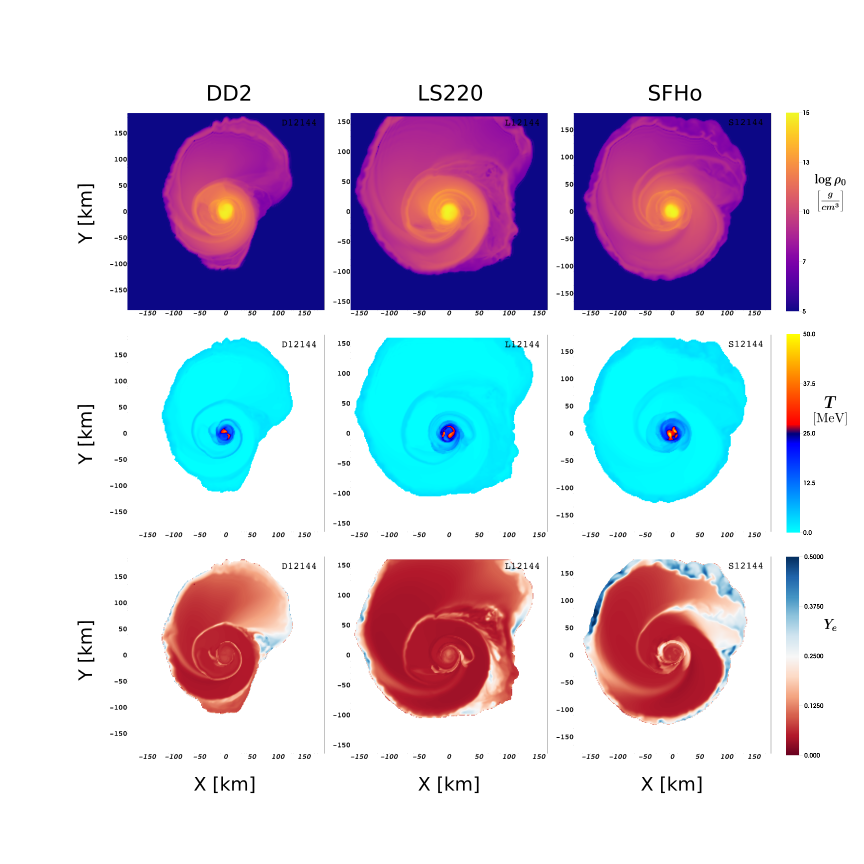}

\caption{
  Density ($\rho_0$), temperature ($T$), and electron fraction ($Y_e$) at 3 ms post-merger for the $1.2M_\odot + 1.44M_\odot$ models. For the temperature and electron fraction plots we threshold on densities above $\sim 10^7 g/cm^3$ to remove the atmosphere, points below this threshold are colored white. At this time, DD2 and LS220 have much more defined tidal-tails than SFHo, but SFHo has a hotter core and higher $Y_e$ in many regions.
}
\label{fig:rho_temp_ye_12144_3ms}
\end{figure*}

Differences in the EOS can have a large impact on the evolution and emission of the system. With the SFHo equation of state, i.e. for the most compact neutron stars, a compact core forms rapidly. In the higher total mass models, the SFHo star collapses promptly to a BH, within a few ms. For the other equations of state (LS220, DD2), more strongly developed tidal features appear at merger. Figure~\ref{fig:rho_temp_ye_12144_3ms} shows each of the $1.2M_\odot + 1.44M_\odot$ models across the three EOS at 3 ms post-merger.

\begin{figure*}[!htbp] \includegraphics[width=\textwidth]{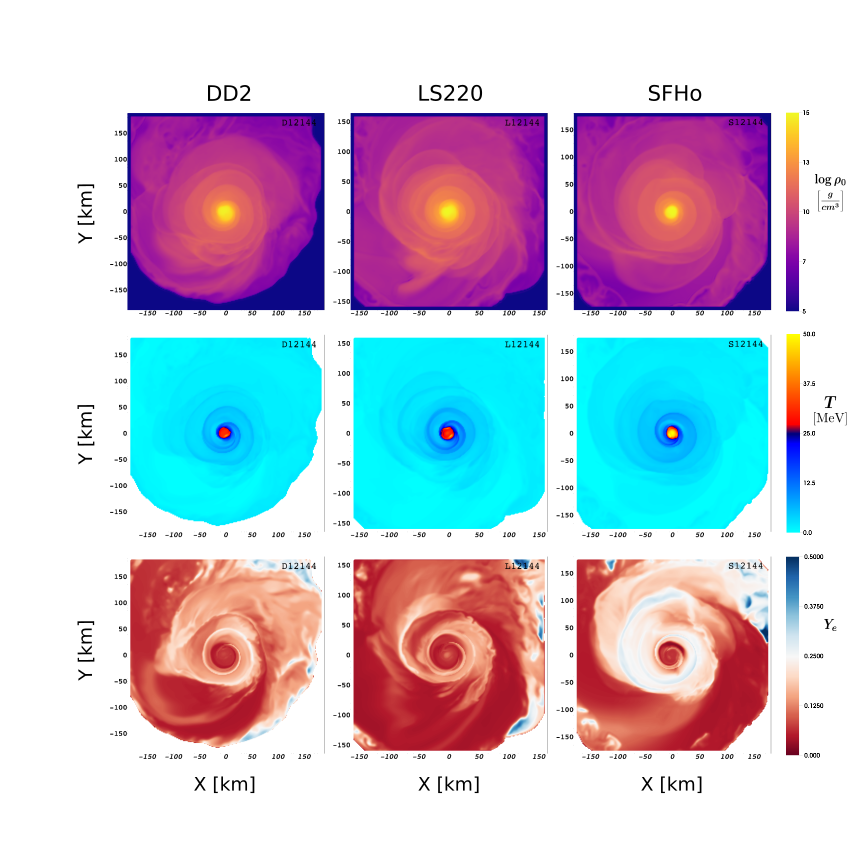}
\caption{
  Density ($\rho_0$), temperature ($T$), and electron fraction ($Y_e$) at 7.5 ms for the $1.2M_\odot+1.44M_\odot$ models. At this time, the tidal-tail and its associated ejecta has left the grid, leaving behind a remnant-core with an accretion torus. SFHo has a hotter core than the other two EOS and a much higher $Y_e$ in the accretion torus.
}
\label{fig:tabular_rho_temp_ye_12132}
\end{figure*}

Finally, at around 7.5 ms after merger, the tidal tail has left the computational grid and the remnant is fully formed with an accretion torus. Figure~\ref{fig:tabular_rho_temp_ye_12132} shows a snapshot of the density profile, temperature profile and electron fraction profile of the binaries at 7.5 ms for the mass-ratio $(1:1.2)$ initial models. As we can see the SFHo EOS has a more dense and hot core than the other two EOS. This is expected as the core bounce for SFHo is much more violent and it occurs deeper in the gravitational potential of the system due to the softness of the EOS. The outflow from the SFHo remnant is also more symmetric because outflows powered by core bounce tend to have a more quasi-spherical morphology than outflows from tidal tails. Stiff EOSs like DD2 undergo less shock heating, but tend to have larger tidal tails. As cold tidal tails are typically neutron-rich, the ejecta in the $z=0$ plane also has a lower $Y_e$.  The electron fraction is also noticeably higher in the disk region for SFHo. This could be due to the fact the SFHo remnant is much hotter and therefore neutrino irradiation effects cause a higher $Y_e$.

\subsection{Matter outflow}

\subsubsection{General properties}

We have compiled a table of the general properties of the ejecta up to either collapse or 7.5 ms post-merger in Table~\ref{tab:matter_ejecta_props}. To help analyze the ejecta properties, we distinguish between two different regions in our grid: polar and equatorial.  The polar ejecta
is defined to be any matter within an angle $\theta < 30^\circ$ of the z-axis (the direction of the total angular momentum of the system), while the equatorial ejecta is the rest of the
ejected matter. From Table~\ref{tab:matter_ejecta_props} we first notice that for each EOS, the majority of the ejecta is equatorial. This is not surprising because tidal disruption tends to produce outflows that remain close to the orbital plane, while the main contribution to the dynamical polar ejecta is from quasi-spherical matter ejection during core-bounce and subsequent oscillations of the post-merger remnant. Given that for spherical mass ejection only $\sim 13\%$ of the matter has $\theta < 30^\circ$, even that quasi-spherical ejecta contributes more to the mass of equatorial ejecta than to the polar ejecta. A more accurate analysis of this table is thus that simulations with the DD2 EOS and mass ratio $(1:1,1:1.1)$, as well as the $(1:1.1)$ simulation with the LS220 EOS have $(10-12)\%$ of their ejecta at $\theta < 30^\circ$ and are thus nearly compatible with spherically symmetric mass ejection. The other configurations have a clear excess of equatorial ejecta, particularly the more asymmetric binaries with the stiffest (DD2) EOS, which have the strongest tidal features.

Secondly, Table~\ref{tab:matter_ejecta_props} shows that the LS220 and SFHo $1.44M_\odot + 1.44M_\odot$ models, which collapse promptly after the merger, have little to no ejecta. Third, the softest EOS, SFHo, has the most ejecta if the remnant avoids prompt collapse ($\sim 0.02M_\odot$), while the LS220 models eject, on average, less mass than the DD2 models. Since LS220 is considered to be softer than DD2, it may be a bit surprising that it ejects less matter. However, as the LS220 EOS has a lower maximum mass than the DD2 EOS, two of the LS220 models lead to relatively rapid collapse of the post-merger remnant to a black hole, which is known to suppress mass ejection. Not all results fit a clear and understandable narrative, however: none of these arguments explain the low mass ejection of the non-collapsing model L12132, or the extremely high mass ejection of the collapsing model S12156. We also note that the dependence of the ejected mass in the mass ratio is clearly non-monotonous. This was also observed by Sekiguchi et al. \cite{sekiguchi2016dynamical} for their SFHo models (but not their DD2 models).
Our results thus add to the growing evidence that while overall trends in mass ejection can be associated with the compactness of neutron stars, the mass asymmetry of the binary, the total mass of the system, and the maximum mass of isolated neutron stars, there is also a significant scatter around 'best fit' models that cannot be easily understood at this point (see also~\cite{Dietrich:2016fpt,radice2018binary}).

Turning to the electron fraction and velocity, we see that $\left<Y_e\right>$ and
$\left<v_\infty\right>$ are much higher in the polar regions and decrease roughly with mass ratio. $\left<Y_e\right>$ tends to hover around $\approx 0.2$ while $\left<v_\infty\right>$ varies between $0.2c$ and $0.3c$. These trends will be touched on in the coming section, where we look at the $Y_e$ and $v_\infty$ distributions. 
\begin{table*}\centering
\newcommand{\ra}[1]{\renewcommand{\arraystretch}{#1}}
\ra{1}
\begin{adjustbox}{width={\textwidth},totalheight={\textheight},keepaspectratio}
\begin{tabular}{lccccccccccc}\toprule
& \multicolumn{3}{c}{Polar $(\theta < 30^\circ)$} & \phantom{abc}& \multicolumn{3}{c}{Equatorial $(\theta > 30^\circ)$} &
  \phantom{abc} & \multicolumn{3}{c}{Total}\\ \cmidrule{2-4}
\cmidrule{6-8} \cmidrule{10-12}
  model\quad\quad\quad\quad &  $M_{ej} (10^{-2}M_\odot)$ & $\left<Y_e\right>$ & $\left<v_\infty\right>$ && $M_{ej} (10^{-2}M_\odot)$ & $\left<Y_e\right>$ & $\left<v_\infty\right>$ && $M_{ej} (10^{-2}M_\odot)$ & $\left<Y_e\right>$ & $\left<v_\infty\right>$\\ \midrule
D12132 & 0.054 & 0.392 & 0.392 && 0.406 & 0.186 & 0.258 && 0.460 & 0.210 & 0.273\\
D12144 & 0.016 & 0.309 & 0.327 && 0.319 & 0.157 & 0.198 && 0.335 & 0.164 & 0.204\\
D12156 & 0.010 & 0.342 & 0.365 && 0.463 & 0.179 & 0.161 && 0.473 & 0.182 & 0.165\\
D144144 & 0.036 & 0.351 & 0.385 && 0.324 & 0.201 & 0.254 && 0.360 & 0.216 & 0.267\\
\midrule
L12132 & 0.011 & 0.331 & 0.377 && 0.083 & 0.188 & 0.204 && 0.094 & 0.205 & 0.224\\
L12144 & 0.026 & 0.292 & 0.345 && 0.357 & 0.186 & 0.185 && 0.384 & 0.194 & 0.196\\
L12156 & 0.004 & 0.302 & 0.336 && 0.230 & 0.190 & 0.131 && 0.234 & 0.192 & 0.135\\
L144144 & 0.000 & 0.344 & 0.362 && 0.012 & 0.213 & 0.255 && 0.012 & 0.217 & 0.258\\
\midrule
S12132 & 0.114 & 0.358 & 0.346 && 1.461 & 0.214 & 0.221 && 1.574 & 0.224 & 0.230\\
S12144 & 0.061 & 0.317 & 0.319 && 0.778 & 0.197 & 0.212 && 0.839 & 0.206 & 0.220\\
S12156 & 0.097 & 0.297 & 0.301 && 1.704 & 0.198 & 0.175 && 1.802 & 0.204 & 0.181\\
S144144 & 0.000 & 0.000 & 0.000 && 0.000 & 0.000 & 0.000 && 0.000 & 0.000 & 0.000\\
\bottomrule
\end{tabular}
\end{adjustbox}
\caption{The first two wide-columns provide the mass ($M_{ej}$), average electron fraction ($\langle Y_e \rangle$) and average asymptotic velocity ($\langle v_\infty \rangle$) of the matter labelled unbound from time t=0 up to 7.5 ms post-merger (or collapse) in the polar and equatorial regions. The last wide-column gives the mass ($M_{ej}$), average electron fraction ($\langle Y_e \rangle$) and average asymptotic velocity ($\langle v_\infty \rangle$) over all regions.}
\label{tab:matter_ejecta_props}
\end{table*}

\begin{figure}[!htbp]
\includegraphics[width=.47\textwidth,trim=200 200 200 100,clip=true]{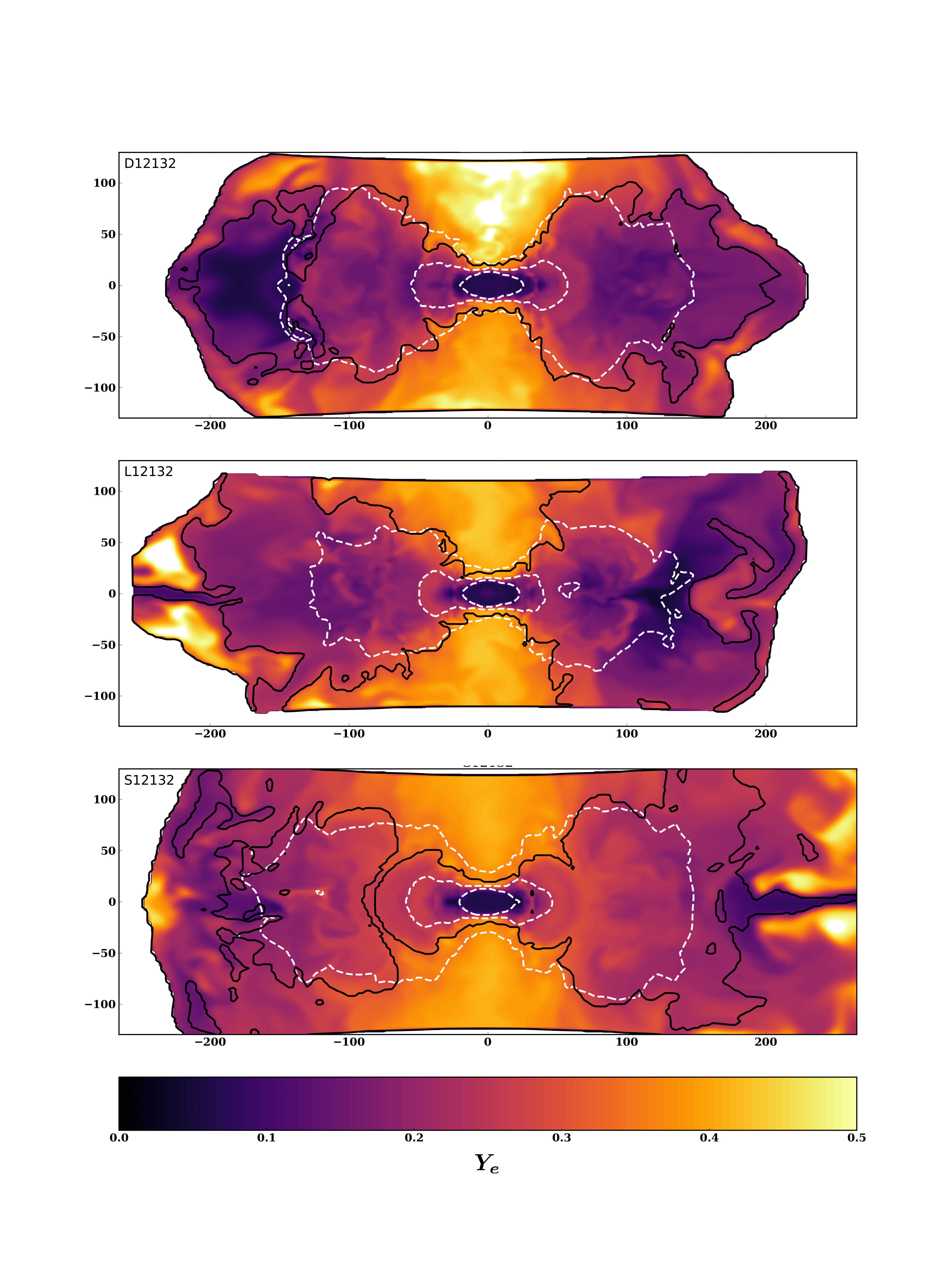}
\caption{
  The electron fraction $Y_e$ for the $1.2M_\odot + 1.32M_\odot$ models at 7.5 ms post merger. Matter that is flagged unbound is outside of the black contours, e.g. the matter in the polar regions at the top and bottom of the grid. The white contours encapsulate material that is above densities
$10^{9}, 10^{11}, 10^{13} g/cm^3$ respectively, with the higher density contours appearing closer to remnant core.
}
\label{fig:ye_table_eos_12132}
\end{figure}

Table~\ref{tab:matter_ejecta_props} quantifies the unbound matter both on the grid (greater than $50M_\odot$ from the remnant) and off the grid. The material on the grid that is still labelled unbound at 7.5 ms is most likely due to neutrino-driven wind. To illustrate that matter ejection continues late in our simulations, we show in Figure~\ref{fig:ye_table_eos_12132}  the $1.2M_\odot + 1.32M_\odot$ models for all EOSs. The black contours separate the bound and
unbound matter, while the white lines are isocontours of the density. Lastly, the colormap denotes the electron fraction, $Y_e$. We note that there is still a small amount of matter labelled unbound, particularly in the SFHo models. Also we again note that the accretion disk is more proton-rich for the SFHo EOS than for the other two systems. Both effects are natural consequences of the hotter post-merger remnant for the SFHo equations of state, which leads to higher neutrino luminosities, stronger neutrino-driven winds, and higher neutrino irradiation of the disk.

\subsubsection{$Y_e$ and $v_\infty$ distributions}

The electron fraction and ejecta velocity are important in determining the attributes of the kilonova and the outcome of r-process nucleosynthesis in the outflows. Indeed, the ejecta velocity has a significant impact on the evolution timescale and brightness of the electromagnetic emission~\cite{2013apj...775...18b}, while the electron fraction is maybe the most important parameter in determining which elements are produced by the r-process. Roughly speaking, for $Y_e\lesssim 0.25$ nucleosynthesis results in the production of heavy r-process elements (mass number $A\gtrsim 120$), while less neutron rich ejecta produces lower mass elements~\cite{lippuner2015}. This has consequences beyond astrophysical nucleosynthesis: in neutron-rich environments, r-process nucleosynthesis produces high-opacity lanthanides and actinides. A higher opacity means that the thermal emission from the ejecta only becomes observable at later times, when the ejecta is cooler. Accordingly, a neutron-rich ejecta typically produces an infrared, week-long kilonova, while a neutron-poor ejecta leads to an optical, day long kilonova~\cite{kasen:2013xka,2013apj...775...18b}.

\begin{figure}[!htbp]
  \includegraphics[width=.5\textwidth]{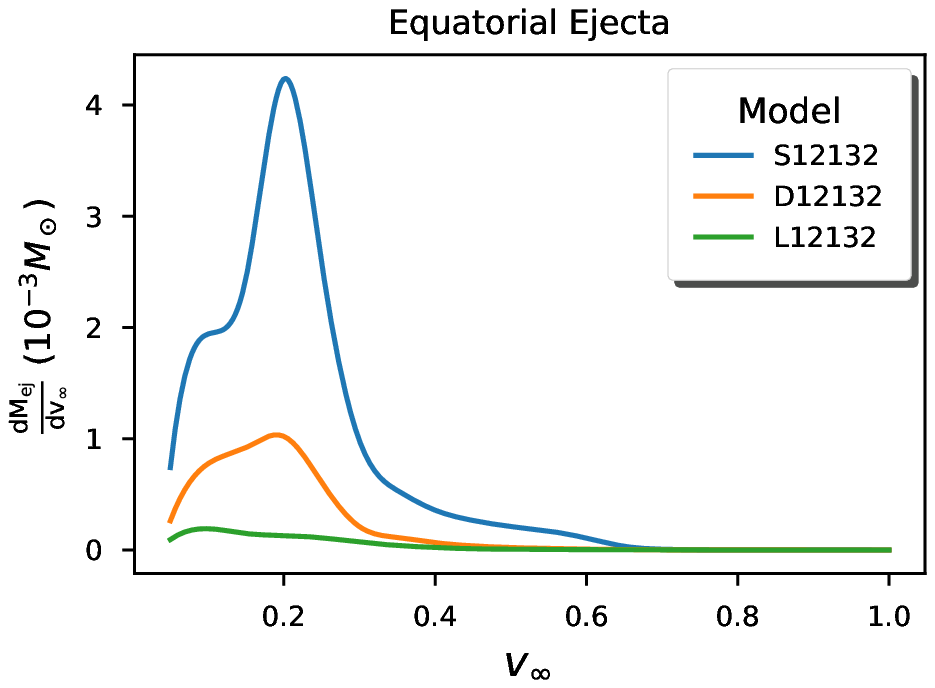}\\
  \includegraphics[width=.5\textwidth]{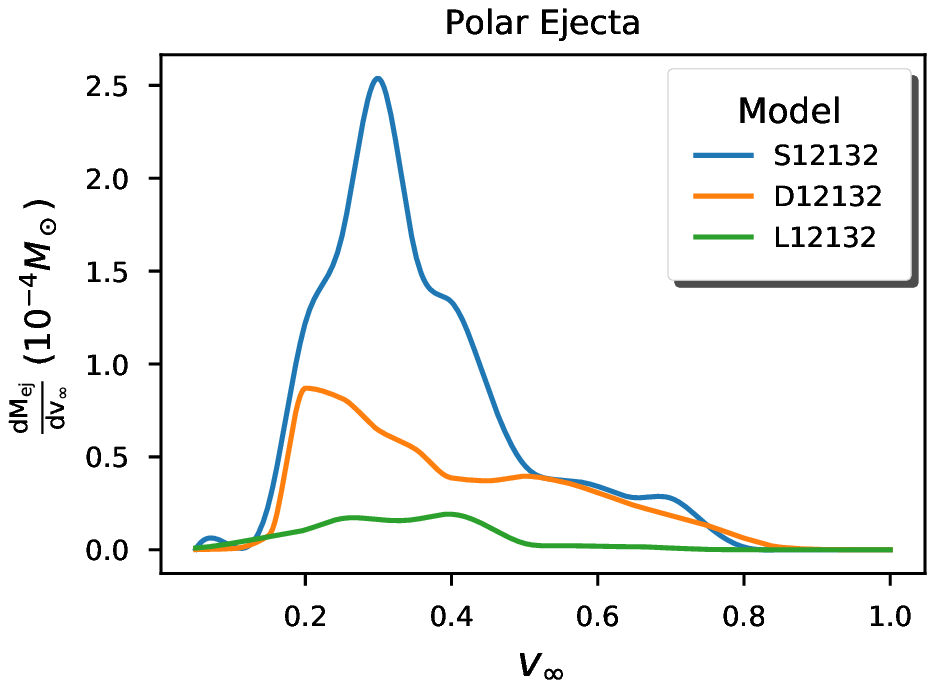}
\caption{
   $v_\infty$ distribution of the polar and equatorial ejecta across different EOS for the $1.2M_\odot + 1.32M_\odot$ models.
}
\label{fig:vinf_ejecta_tab_eos}
\end{figure}

Figure~\ref{fig:vinf_ejecta_tab_eos} shows the inferred velocity of the ejecta at infinity across EOS for both the equatorial and polar regions at 7.5 ms post-merger. First we notice that the polar ejecta generally have a larger velocity than the equatorial ejecta. This is most likely due to the polar ejecta originating primarily from core-bounce and the collision of the two neutron stars, which produce faster moving ejecta than the more equatorial-plane bounded tidal ejecta. Furthermore, the faster-moving core-bounce ejecta traveling in the equatorial-plane may collide with tidal ejecta and be slowed down. 
We also notice that overall, most of the ejecta is at the lower end of the velocity range, $v_\infty \sim 0.2c$ with very little ejecta at $v_\infty > 0.8c$ for any of the models pictured here. LS220 appears to eject matter with smaller velocities, although not very conclusively considering the small amount of matter ejected by that simulation.
Thus, across EOSs, we cannot establish any robust differences in velocity distribution.

\begin{figure}[!htbp]
  \includegraphics[width=.47\textwidth]{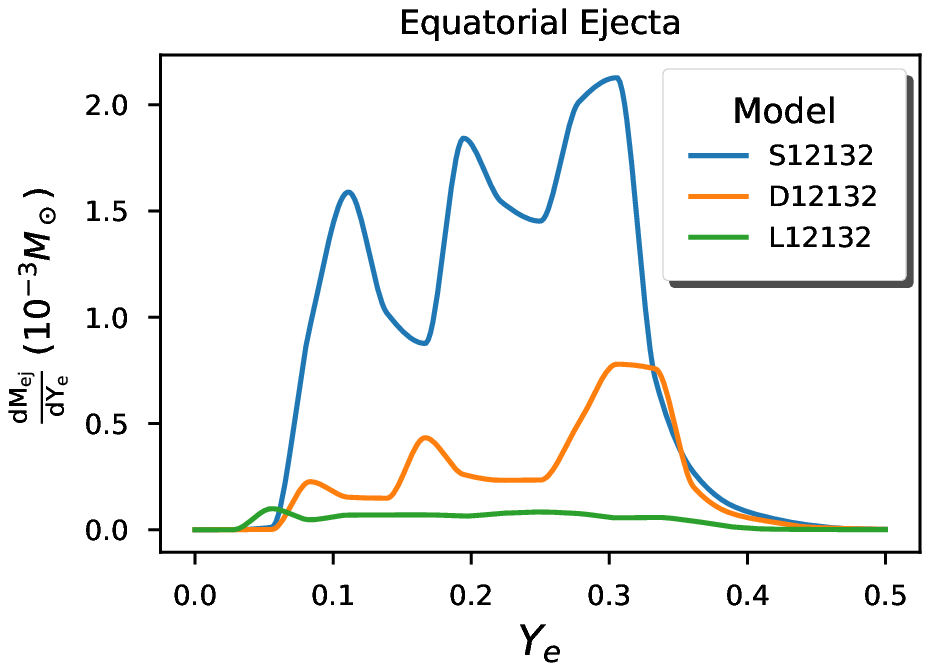}\\
  \includegraphics[width=.47\textwidth]{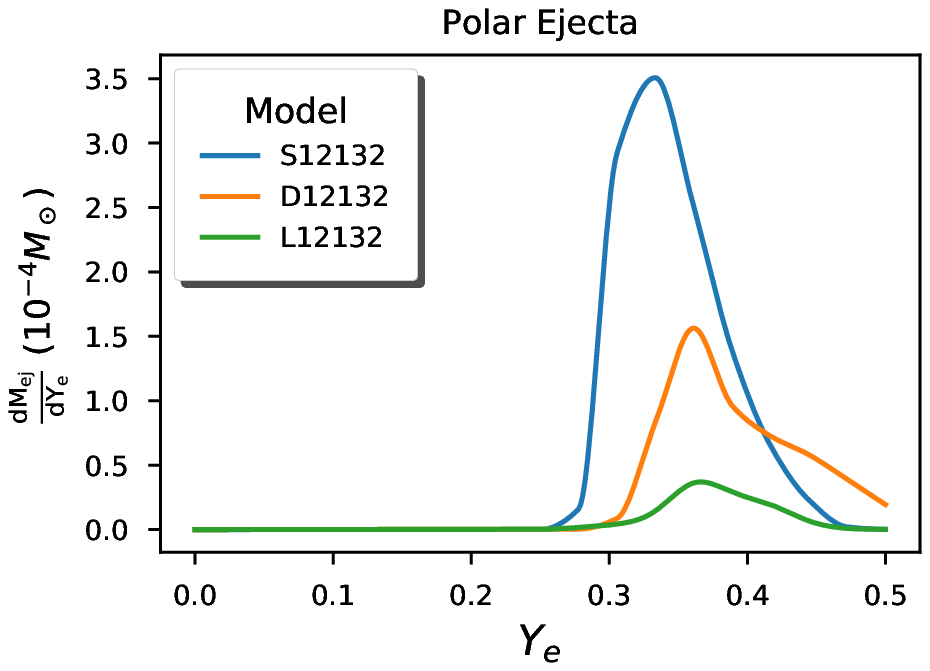}
\caption{
   $Y_e$ distribution of the polar and equatorial ejecta across different EOS for the $1.2M_\odot + 1.32M_\odot$ models .
}
\label{fig:ye_ejecta_tab_eos}
\end{figure}

  Figure~\ref{fig:ye_ejecta_tab_eos} shows the electron fraction of ejected material for different EOSs in both the equatorial and polar regions. Qualitatively, the shapes of the distributions appear mostly EOS-independent.
  The equatorial ejecta have a broad range of electron fraction values whereas the polar ejecta, which make a much smaller fraction of the ejected mass, take on much higher $Y_e$ values on average. The production of dynamical ejecta with a broad $Y_e$ distribution and the presence of higher-$Y_e$ outflows in the polar regions are both robust results observed in all simulations of binary neutron star mergers using advanced neutrino transport methods~\cite{wanajo2014,sekiguchi2016dynamical,foucart:2015gaa,foucart2016impact}.
  
\begin{figure}[!htbp]
  \includegraphics[width=.47\textwidth]{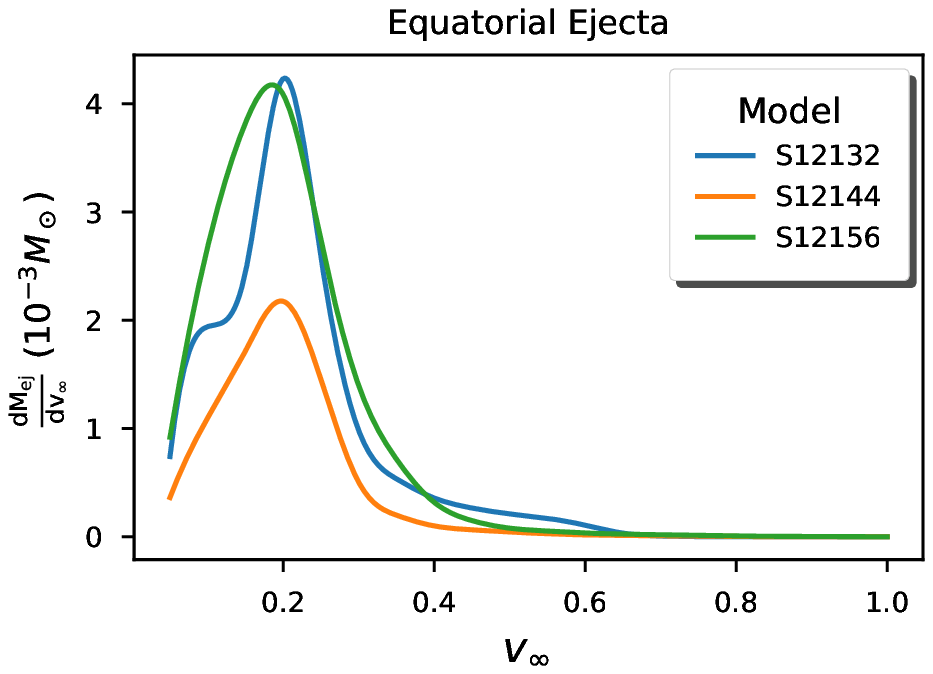}\\%
  \includegraphics[width=.47\textwidth]{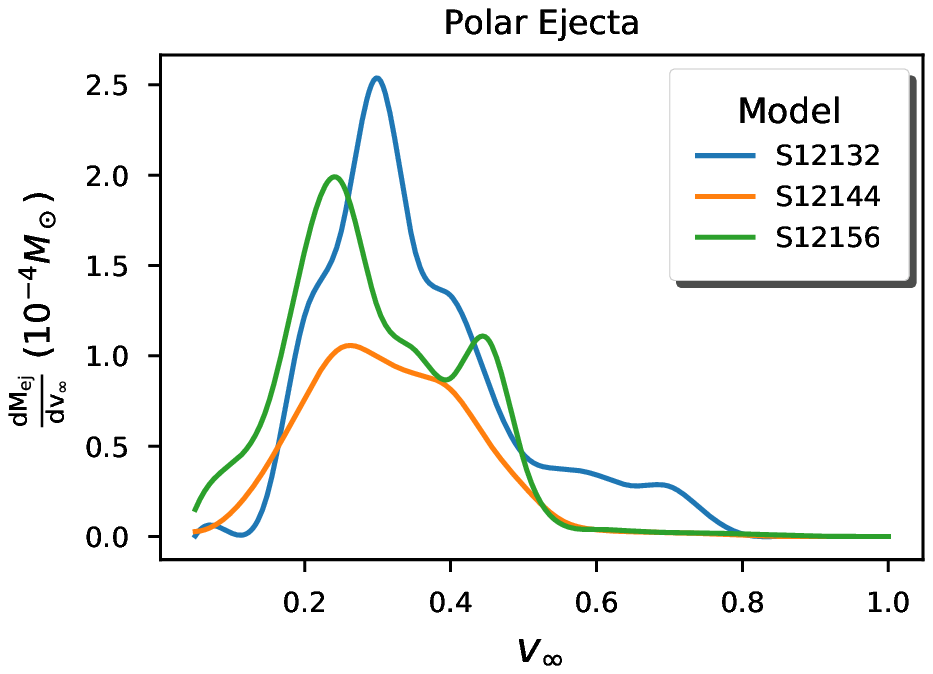}
\caption{
   $v_\infty$ distribution of the polar and equatorial ejecta across different mass ratios for the SFHo EOS.
}
\label{fig:vinf_ejecta_tab_sfho}
\end{figure}

Figure ~\ref{fig:vinf_ejecta_tab_eos} and ~\ref{fig:ye_ejecta_tab_eos} investigated EOS-dependence at fixed component-masses. We now consider varying the component masses, while keeping the EOS fixed. Figure~\ref{fig:vinf_ejecta_tab_sfho} shows the distribution of $v_\infty$ 
for the SFHo EOS at different mass ratios.
As the system becomes more asymmetric, the average velocity of the outflows generally decreases, and the tail of the distribution at high velocities is suppressed. The lack of high-velocity ejecta for more asymmetric binaries could affect the ability of these systems to power fast, blue/UV emission~\cite{metzger2017}. The decrease in the average velocity of the ejecta as the binary becomes more asymmetric is notable, as that trends is exactly the opposite of what is observed in black hole-neutron star mergers~\cite{kawaguchi2016models,foucart2016dynamical}. We interpret this difference as being due to the impact of the core-bounce ejecta, which is absent in black hole-neutron star mergers: while the velocity of the tidal ejecta may increase with the asymmetry of the system, the relative importance of the (faster) core-bounce ejecta decreases for more asymmetric systems, thus reducing the average velocity of the ejecta for binary neutron star systems.

\begin{figure}[!htbp]
     \includegraphics[width=.495\textwidth]{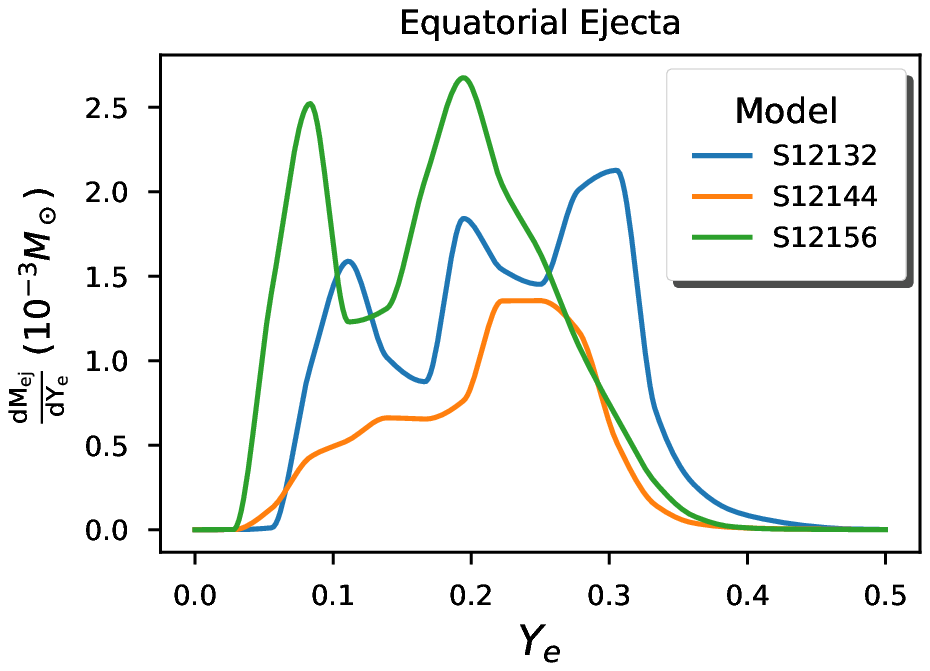}\\
 \includegraphics[width=.495\textwidth]{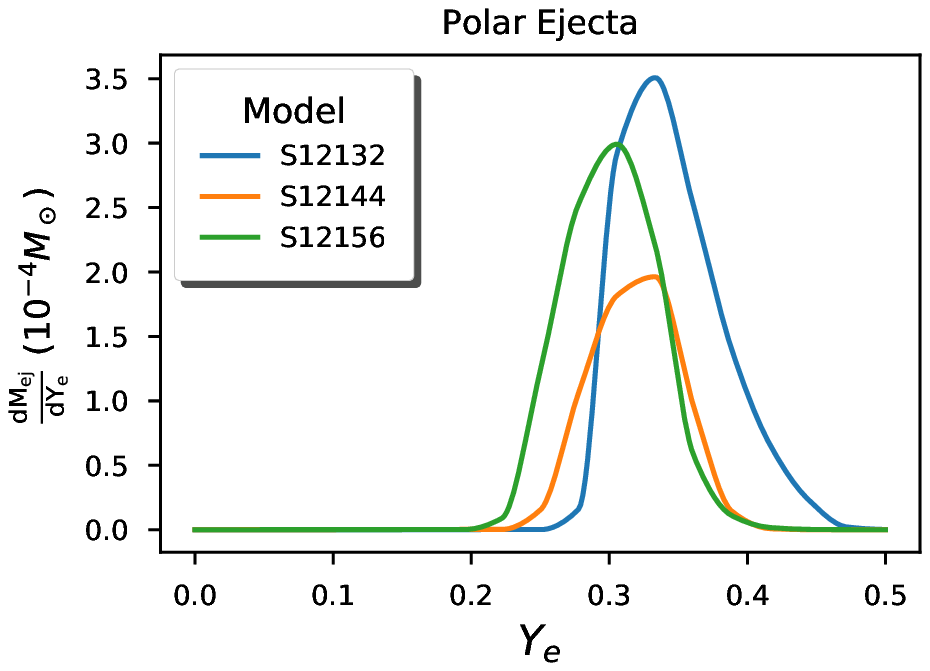}
\caption{
   $Y_e$ distribution of the polar and equatorial ejecta across different mass ratios of the SFHo EOS.
}
\label{fig:ye_ejecta_tab_sfho}
\end{figure}

Finally, Figure~\ref{fig:ye_ejecta_tab_sfho} shows the distribution of electron fraction for the SFHo EOS at different mass ratios. We see that as mass asymmetry increases, the average $Y_e$ of the ejecta decreases.
The distributions remain broad, with higher values in the polar regions.

\subsubsection{Long-term Emission}
\label{sec:long_term}
\begin{figure}[!htbp]
\includegraphics[width=.48\textwidth]{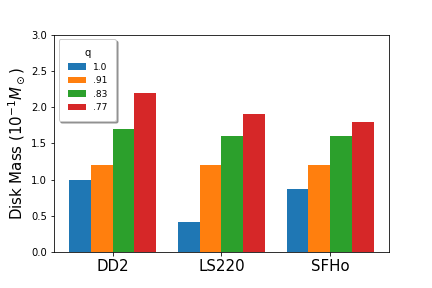}
\caption{
  Disk masses at 7.5 ms post-merger or just before collapse if remnant is a black hole.
}
\label{fig:disk_masses}
\end{figure}

Analysis of the kilonova emission associated with GW170817 showed that the ejecta mass was on the order of $\sim 0.05M_\odot$ \cite{Metzger:2017wot,shibata2017gw170817}. This is above the estimates from dynamical ejecta seen in Table~\ref{tab:matter_ejecta_props}, or for that matter above the mass of dynamical ejecta in any low-eccentricity, low-spin neutron star merger simulation performed to date (see e.g.~\cite{Dietrich:2016fpt} for a summary). This is not really a problem, however: our analysis ignores other forms of ejecta that become more important over longer evolution time scales. 
After merger, magnetically-driven~\cite{siegel:2017nub,fernandez2019long} and neutrino-driven~\cite{just2014} winds as well as outflows due to viscous angular momentum transport and $\alpha$-particle recombination~\cite{fernandez2013} should combine to unbind a large fraction of the remnant disk ($\sim 40\%$ for disks around black holes in~\cite{fernandez2019long}). Based on these results for the long-term evolution of post-merger disk, we can provide order-of-magnitude estimates of the evolution of the remnant produced at the end of our simulations. We consider
the following definition of the disk mass:

%%%%%
\begin{equation}
  \label{eqn:disk_mass}
M_{\text{disk}} = \int_{\rho < 10^{13}g/cm^3} \rho_0 W \sqrt{\gamma} \mathrm{d}^3x,
\end{equation}

which has been used in previous works, see \cite{radice2018binary,shibata2017gw170817}. This definition is used for two reasons. Firstly, based on visualizations of the remnant, such as Fig.~\ref{fig:ye_table_eos_12132} one can clearly see that the $10^{13}g/cm^3$ density contour encapsulates only the core of the remnant. Secondly, it was shown that for densities approximately below $10^{13}g/cm^3$, the matter is rotationally bound with roughly a $r^{-3/2}$ fall-off\cite{hanauske2017rotational}. Unfortunately $M_{disk}$ defined by Eq.~(\ref{eqn:disk_mass}) drifts to larger values over time and indeed it has not converged yet at 7.5 ms. Nonetheless, we note similar evolution of this quantity over time for all EOS and mass-ratios, so using estimates at 7.5 ms will give us a crude estimate on the lower bound of the disk mass.  We compute the disk masses at 7.5-ms post-merger except for the models that collapse  (e.g. S12156, S144144, L12156, L144144) where we compute the disk mass right before collapse. The masses are summarized in Fig.~\ref{fig:disk_masses}. 

We first note that the stiffer EOS produce larger disks. Moreover, disk mass increases as the degree of binary asymmetry increases. Most importantly, all but one simulation (L144144) can potentially eject enough mass to explain the kilonova observed after GW170917, with required efficiency for the conversion of disk mass to disk outflows ranging from $\sim 25\%$ to $\sim 50\%$. This is well within the (very uncertain) range of efficiencies allowed by current post-merger evolutions.

\subsection{Neutrino Emission}
Finally we turn to the neutrino emission of our models. Neutrinos play an important role in the post-merger evolution, not only cooling the central massive regions, but also transporting energy into the surface layers of the disk, providing a long-term outflow. On top of this, neutrinos deposit net lepton number through weak interactions which can drastically change the electron fraction of outflows and thus affect the kilonova emission. Furthermore, they can %shuffle 
deposit large amounts of thermal energy into the polar regions through $\nu\bar\nu$ annihilation, potentially contributing to the creation of baryon free regions above the remnant and helping the production of jet-like emissions.

\begin{figure}[!htbp]
\includegraphics[width=.48\textwidth,trim=0 0 0 0,clip=true]{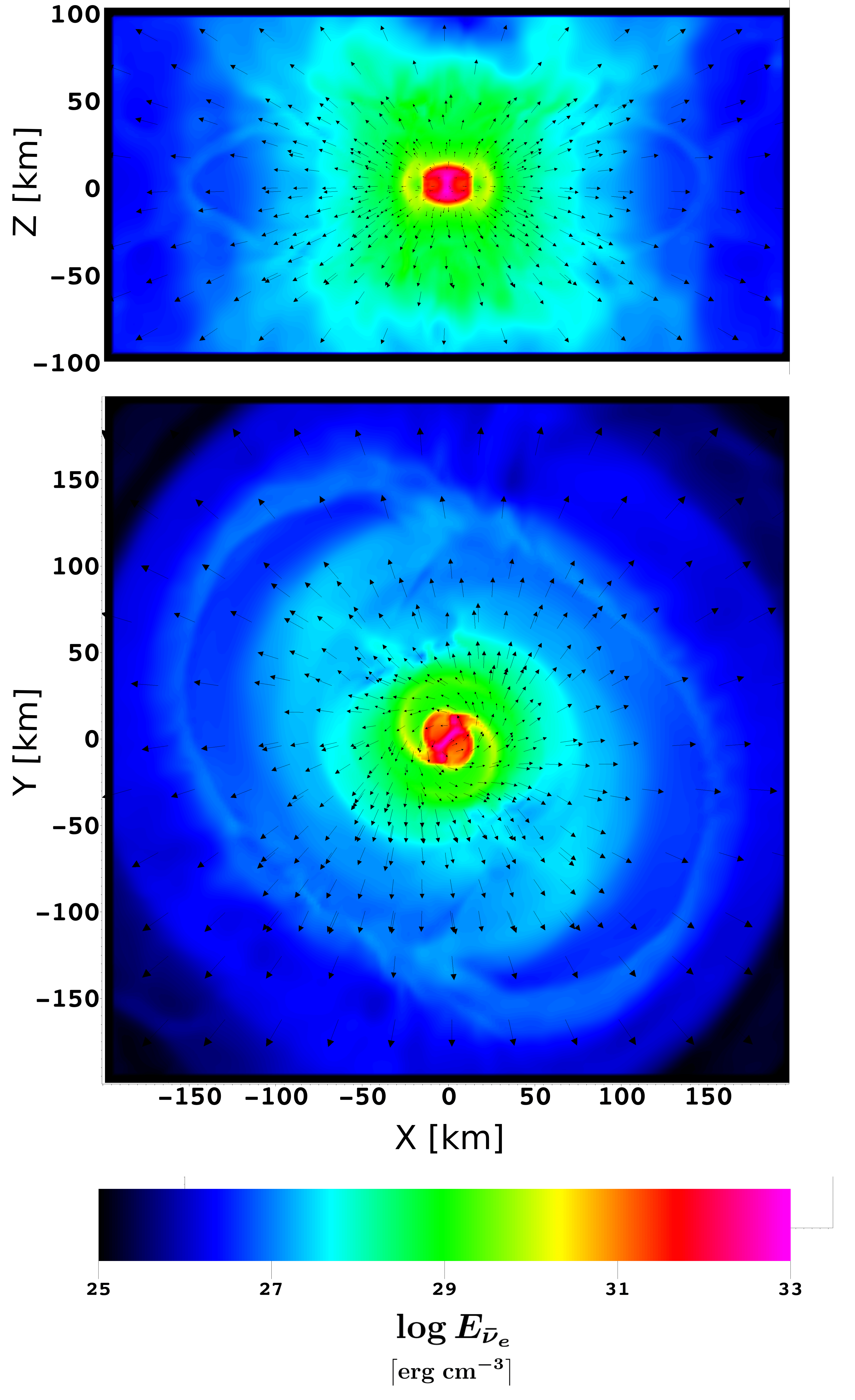}
\caption{
  Anti-electron neutrino energy density (first moment) for the 1.44$M_\odot$ + 1.44$M_\odot$ DD2 model at 3 ms post-merger. \textbf{Top panel:} A xz-slice of the anti-electron neutrino energy density $E_{\bar\nu_e}$ with the arrows showing the effective neutrino transport velocity $v_\nu^i = \alpha \frac{E_{\bar\nu_e}}{F^i_{\bar\nu_e}} - \beta^i$. We see a large energy density near the polar regions of the remnant, where the density is less and the neutrinos are free to stream (While an overdensity of neutrinos in the polar region is expected to be qualitatively correct, we note that the use of the approximate Minerbo closure leads us to overestimate the neutrino density at the poles~\cite{foucart2016impact}). \textbf{Bottom panel:} A xy-slice of the anti-electron neutrino energy density $E_{\bar\nu_e}$ with the arrows showing the effective neutrino transport velocity. We see the neutrinos advecting with the fluid in the dense regions near the core.
}
  \label{fig:d144144_3ms_ENua}
\end{figure}

\begin{figure}[!htbp]
  \includegraphics[width=.45\textwidth]{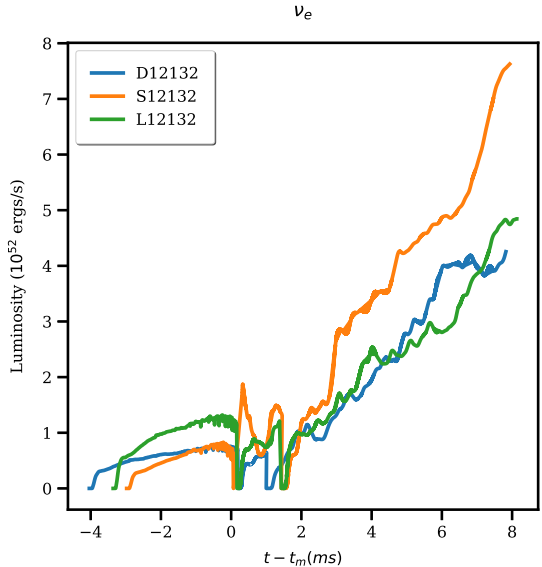}
\caption{
  Electron neutrino luminosity for the 12132 models up to around 7.5 post-merger. We see that for the softer SFHo EOS, there is a larger luminosity. Similar trends hold for the other neutrino species.
}
\label{fig:nulum_table_eos}
\end{figure}

As seen in the bottom portion of Fig.~\ref{fig:d144144_3ms_ENua}, which shows the electron anti-neutrino energy density for the D144144 model at 3 ms post-merger, the main emission regions are the hot, dense parts of the remnant; the central core, and the shocked tidal arms. In these dense parts of the remant the neutrinos are advected with the flow. The main escape route for the neutrinos trapped in the dense part of the remnant is via the polar regions, which show a greater neutrino energy density in the top portion of Fig.~\ref{fig:d144144_3ms_ENua} than the equatorial parts. Due to the more compact and hotter remnants, we would expect that the SFHo models might show greater neutrino luminosity than the other EOS and this is confirmed in Figure~\ref{fig:nulum_table_eos} for the electron neutrino luminosity. Although we do not plot them, the other species show exactly the same type of behavior, with SFHo having a much higher luminosity in all three species evolved in the transport scheme.

\begin{figure*}[!htbp]
  \includegraphics[width=1.2\textwidth,center]{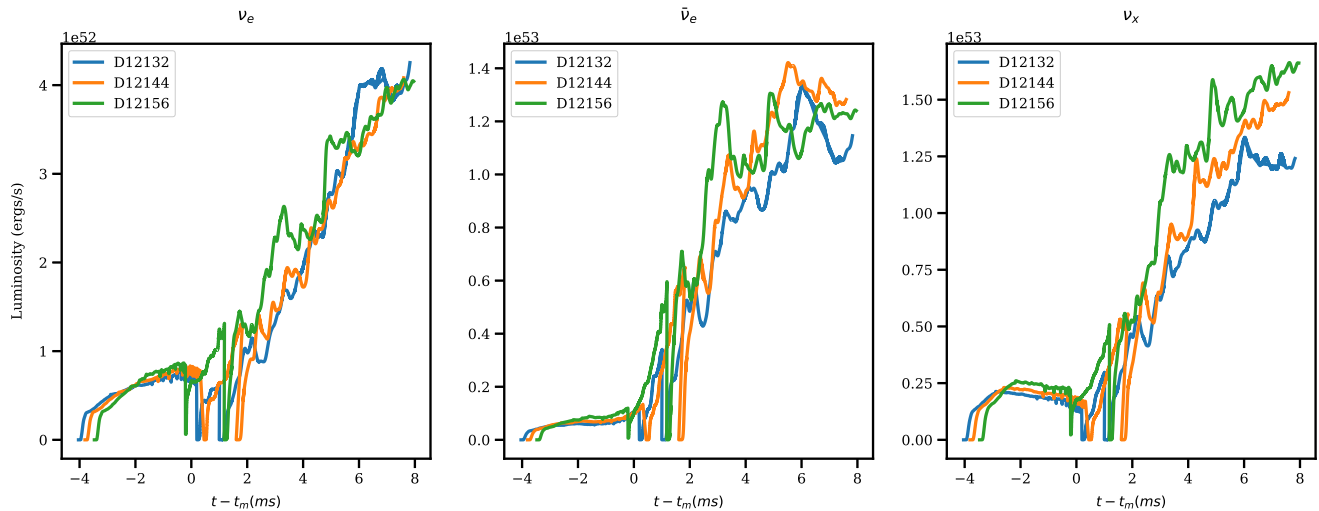}
\caption{
  Neutrino luminosity for DD2 runs up to around 7.5 ms across neutrino species.
}
\label{fig:nulum_table_dd2}
\end{figure*}

As the mass ratio changes, we find only mild variations in the neutrino luminosity as shown in Fig.~\ref{fig:nulum_table_dd2}. The heavy neutrino luminosity has the strongest dependance on the mass ratio, showing a sizable increase as the binary becomes more asymmetric. This is in contrast to Sekiguchi et al. \cite{sekiguchi:2016} which finds a decrease in neutrino luminosity as the binary asymmetry increases. However, Sekiguchi et al. fix the total mass at $2.7M_\odot$. Since we only fix the lower mass neutron star in the binary, a larger mass asymmetry implies a higher total mass, more compact remnants, higher temperatures and higher neutrino emission. 
In our simulations, the impact of the total mass of the system is cancelled out by the change in mass ratio, which, as Sekiguchi showed, tends to decrease the neutrino luminosity of the post-merger remnant. 

\begin{figure*}[!htbp]
\centering \includegraphics[width=\textwidth,trim=300 0 300 0,clip=true]{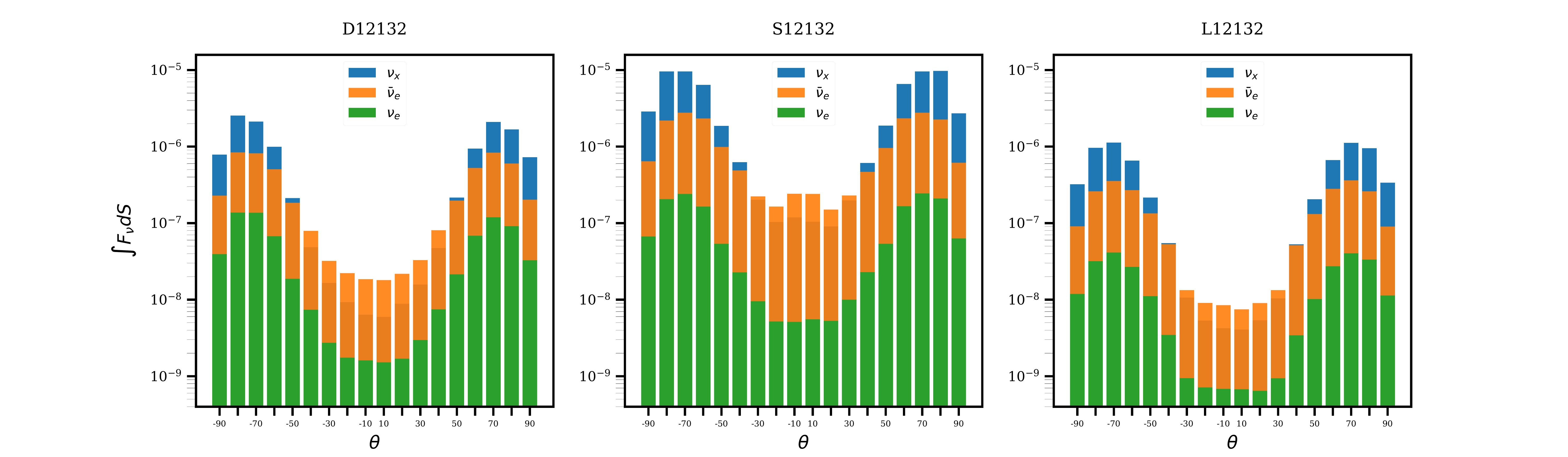}
\caption{
  Neutrino flux density moment as a function of angle for different EOSs at 7.5 ms post-merger. The angle is defined with respect to the equatorial plane with $0^\circ$ being the equator, and $\pm 90^\circ$ being the North and South poles. The different neutrino species are color coded.
}
\label{fig:theta_nu_eos}
\end{figure*}

Next we take a look at the angular distribution of the neutrino emission. Figure~\ref{fig:theta_nu_eos} shows the neutrino flux density as a function of its angle with respect to the equatorial plane at 7.5 ms after merger. From this figure, one can clearly see that most of the neutrinos are emitted in the polar directions. Once a disk forms, the neutrinos are mostly confined within a 40 degrees cone around the poles, with an amplitude peak at 30 - 40 degrees from the poles.
As argues in~\cite{foucart2016impact}, this peak is probably due to neutrinos beamed from the shocked tidal arms, which become less optically thick as time passes. The confinement of the neutrinos to the polar directions stems from the fact that neutrinos escape through the low-density regions above and below the disk and are confined by the optically thick accretion disk. As the EOS changes, the distributions stay qualitatively the same. The binary mass ratio as an equally negligible impact on the angular distribution of neutrinos, and is thus not depicted here.

\begin{figure}
  \includegraphics[width=.45\textwidth]{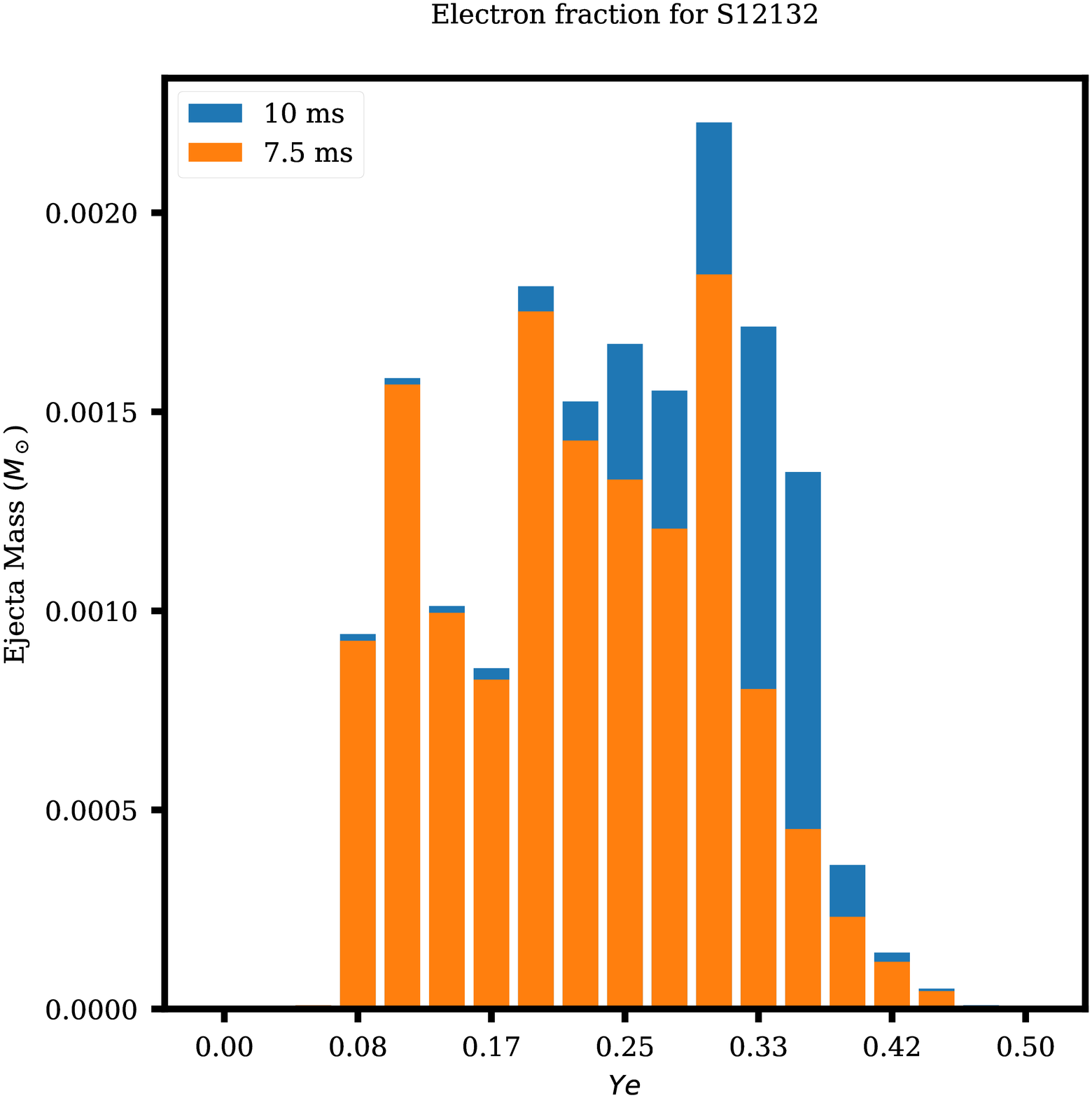}
\caption{
  Electron fraction distribution for the off-grid ejecta of the $1.2M_\odot + 1.32M_\odot$ SFHo model at 7.5 ms and 10 ms post-merger. One can clearly see that more matter is being ejected at higher $Y_e$ fractions due to neutrino irradtiation.
}
\label{fig:s12132_75_and_10}
\end{figure}

While neutrino irradiation of the disk corona can drive outflows, the total mass of these outflows is negligible, especially over the fairly short time scales evolved here.
We do however see some effects of neutrino irradiation in the electron fraction of the outflows. 
Figure~\ref{fig:s12132_75_and_10} shows the electron fraction distributions for all outflows escaping the grid before 7.5-ms and 10-ms post-merger. As we can see, 
the disk outflows observed at the end of the simulation have a particularly high electron fraction.
The reason for this is that in the presence of strong electron neutrino luminosity, the neutrino capture processes are activated, increasing the overall electron fraction.

\section{Conclusion and Future Work}

We presented a first set of SpEC simulations of unequal mass neutron star mergers using 3 nuclear-theory based equations of state (SFHo, LS220, DD2), and a two-moment grey neutrino transport scheme with an improved energy estimate based on evolving the number density. This set of simulations varied the equation of state of the neutron stars and the total mass and mass-ratio of the binary in order to determine any robust trends between these parameters and emission characteristics. We found the following

\begin{itemize}
\item For the models that promptly collapsed, namely L144144 and S144144, there was little to no ejecta and the disk masses were much smaller. 
This is consistent with other simulations of promptly-collapsing equal mass neutron star-neutron star binaries.
\item The softest models, those of the SFHo EOS, had the most ejecta and SFHo was the only EOS with $M_{ej} \sim 10^{-2}M_\odot$. 
\item In a majority of the simulations, we found that matter was preferentially ejected more than $30^\circ$ away from the poles. This result is however not universal: for near equal-mass systems and stiff EOSs, we find that the outflows are dominated by the more spherically symmetric matter ejected at core-bounce and during subsequent oscillations of the neutron stars.
\item The $Y_e$ distributions were very broad $(\sim 0.06 - 0.48)$ and quite similar across EOS and mass-ratio. $\langle Y_e \rangle$ is around $\sim 0.2$, but increases over time as the ejected matter is irradiated by neutrinos. We see a decrease in the average $Y_e$ with increasing binary asymmetry, but there is some scatter, so the relationship might be more complicated.
\item The asymptotic velocity distributions were also very broad $(\sim 0.05c - 0.7c)$, with only a small amount of matter in the high-velocity tail of these distributions and average velocities
  within the range $(0.2 - 0.3)c$.
We see a decrease in the average asymptotic velocity for more asymmetric mass ratios.
\item The disk masses of the models at 7.5 ms post-merger appear to increase with mass-ratio and stiffness of the EOS. The softest EOS, SFHo, has a disk with a much higher $Y_e$ than the other EOS, presumably because softest EOSs are associated with more compact stars, hotter post-merger remnants, and thus stronger neutrino irradiation of both the disk and the matter outflows.
\item Neutrino emission was accordingly largest for the SFHo EOS.
We found no dependence of the neutrino luminosity on the mass-ratio, however, except in the case of heavy-lepton neutrinos. Most of the neutrino emission was in the polar region and we did not find any dependence of its morphology in either the mass-ratio or the EOS.
\end{itemize}

The results presented here are limited by two important assumptions in our simulations. Firstly, the absence of magnetic fields. Over the short time scales considered here, magnetohydrodynamics effects are not expected to affect the evolution of the neutron star remnant, but could drive additional outflows from the disk \cite{kiuchi2014,neilsen2014magnetized}. Over longer time scales, magnetic fields would be critical to the spin evolution of the remnant neutron star, angular momentum transport, heating in the disk, and possibly the formation of relativistic jets and magnetically-driven outflows. The second limitation is the relatively small numerical grid on which we evolve the equations of general relativistic hydrodynamics. The grid limits our ability to measure the mass and properties of the outflows accurately. We expect to address these issues in the future. However, we do not expect these assumptions to significantly affect the main results of this work, i.e. the properties of the ejecta and the trends across EOS and mass-ratio in the first 7.5 ms after merger.

%%%%%%%%%%%%%%%%%%%%%%%%%%%%%%%%%%%%%%%%%%%%%%%%%%%%%%%%%%%%%%%%%%%%%%%%%%%%%%%
% Acknowledgments
%%%%%%%%%%%%%%%%%%%%%%%%%%%%%%%%%%%%%%%%%%%%%%%%%%%%%%%%%%%%%%%%%%%%%%%%%%%%%%%

\begin{acknowledgements}
  We would like to acknowledge helpful discussions with Tim Dietrich, Wyatt Brege and Sergei Ossokine. 
  F.F. gratefully acknowledges
support from NASA through grant 80NSSC18K0565,
and from the NSF through grant PHY-1806278. H.P. gratefully acknowledges support from the
NSERC Canada. L.K. acknowledges support from NSF grant
PHY-1606654. M.S. acknowledges support from NSF Grants
PHY-170212 and PHY-1708213. L.K. and M.S. also thank
the Sherman Fairchild Foundation for their support. Computations were performed on the supercomputer Briaree from the Universite de Montreal, managed by Calcul Quebec and Compute Canada. The operation of these supercomputers is funded by the Canada Foundation for Innovation (CFI), NanoQuebec, RMGA and the Fonds de recherche du Quebec - Nature et
Technologie (FRQ-NT).  Computations were also
performed on the Minerva cluster at the Max-Planck-Institute for
Gravitational Physics, and the GPC and Niagara supercomputers at the
SciNet HPC Consortium~\cite{scinet}. SciNet is funded by: the Canada
Foundation for Innovation; the Government of Ontario; Ontario Research
Fund - Research Excellence; and the University of Toronto.
\end{acknowledgements}
%
%%%%%%%%%%%%%%%%%%%%%%%%%%%%%%%%%%%%%%%%%%%%%%%%%%%%%%%%%%%%%%%%%%%%%%%%%%%%%%%
\section*{References}
%%%%%%%%%%%%%%%%%%%%%%%%%%%%%%%%%%%%%%%%%%%%%%%%%%%%%%%%%%%%%%%%%%%%%%%%%%%%%%%
\bibliography{References_final}
\appendix
\end{document}